\documentclass[preprint,aps ,nofootinbib]{revtex4}
\usepackage{graphicx}
\usepackage{epsfig}
\usepackage{amsmath}
\usepackage{amsfonts}
\usepackage{amssymb}
\usepackage{color}
\usepackage{dcolumn}
\usepackage{hyperref}
\usepackage{multirow,gensymb}
\usepackage{booktabs}
\usepackage{tikz}
\usetikzlibrary{calc}

\providecommand{\U}[1]{\protect\rule{.1in}{.1in}}

\usepackage{soul}
\newcommand{\f}{\begin{equation}}
\newcommand{\ff}{\end{equation}}
\newcommand{\fa}{\begin{eqnarray}}
\newcommand{\ffa}{\end{eqnarray}}

\begin{document}
\title{The holographic Fermions over the ionic lattice with CDW}
\author{Kai Li $^{1,2}$}
\email{lik@ihep.ac.cn}
\author{Yi Ling $^{1,2}$}
\email{lingy@ihep.ac.cn}
\author{Peng Liu $^{3}$}
\email{phylp@email.jnu.edu.cn}
\author{Chao Niu$^{3}$}
\email{niuchaophy@gmail.com (Corresponding author)}
\author{Meng-He Wu$^{4}$}
\email{mhwu@njtc.edu.cn (Corresponding author)}

\affiliation{%  
    $^1$ Institute of High Energy Physics, Chinese Academy of Sciences, Beijing 100049, China;\\
    $^2$ School of Physics, University of Chinese Academy of Sciences, Beijing 100049, China;\\
    $^3$ Department of Physics and Siyuan Laboratory, Jinan University, Guangzhou 510632, China;\\
     $^4$ College of Physics and Electronic Information Engineering, Neijiang Normal University, Neijiang 641112, China.
}

\begin{abstract}
We study the holographic Fermion as a probe over the background with ionic lattice, which may undergo a phase transition with the development of charge density wave by the spontaneous breaking of the translational symmetry. We focus on the structure of the Fermi surface within different Brillouin zones and demonstrate how the presence of CDW in the background affects the formation of the band gap in the momentum space. Specifically, we find the formation of the CDW enhances the amplitude of spectral function as well as the momentum of the Fermi surface. Furthermore, we are concerned with the change of the Fermi surface with the doping parameter as well as the lattice amplitude. Interestingly, we find that the radius of the Fermi surface expands with the increase of the doping parameter and finally may cross the first Brillouin zone. Additionally, the width of band gap becomes larger with the increase of the lattice amplitude as well, which is consistent with the observation in condensed matter experiments.

\end{abstract}

\maketitle

\section{Introduction}

In the past decade the advent of holographic techniques, rooted in the AdS/CFT correspondence \cite{Maldacena:1997re,Gubser:1998bc,Witten:1998qj}, has opened unprecedented avenues for studying strongly correlated systems in condensed matter physics, where the traditional perturbative method usually fails \cite{Hartnoll:2009sz,Iqbal:2011ae,Hartnoll:2016apf,Zaanen:2015oix,Ammon:2015wua}. On the one hand, diverse holographic models with abundant structure of bulk geometry have been constructed to simulate the fascinating phenomena in condensed matter physics, such as the high-temperature superconductivity \cite{Gubser:2008px,Hartnoll:2008vx,Hartnoll:2008kx}, strange metal \cite{Hartnoll:2009ns,Faulkner:2010zz} as well as quantum critical phenomenon \cite{Cubrovic:2009ye}. Among these, systems featuring lattices \cite{Hartnoll:2012rj,Horowitz:2012ky,Horowitz:2012gs, 
 Horowitz:2013jaa,Ling:2013nxa,Andrade:2013gsa} and charge density waves (CDW) \cite{Donos:2011bh,Withers:2013loa,Donos:2013gda,Donos:2013wia,Withers:2013kva,Ling:2014saa} represent paradigmatic examples of materials where the translational symmetry is broken explicitly or spontaneously, respectively. Later, the interplay between the lattice and CDW has been developed in both homogeneous lattice models and inhomogeneous lattice models \cite{Andrade:2015iyf,Baggioli:2016oju,Jokela:2017ltu,Krikun:2017cyw,Andrade:2017leb,Andrade:2017ghg,Ammon:2019wci,Andrade:2020hpu,Baggioli:2021xuv,Baggioli:2022pyb}. In this direction the advantage of inhomogeneous lattices has been revealed in recent study on the holographic construction of Mott insulator \cite{Andrade:2017ghg}. It indicates that the lock-in effect of the commensurate state can be implemented only on the inhomogeneous lattices, rather than homogeneous lattices \cite{Andrade:2015iyf,Andrade:2017leb}. This feature plays an essential role in understanding the phase structure of the high-temperature superconductivity. It is also based on this feature that the holographic description of the striped superconductor is realized only recently in \cite{Ling:2020qdd, Li:2024ybq}, where the pair density wave is emerged explicitly as the intertwined phase of charge density waves (CDW) order and the superconducting (SC) order. On the other hand, exploring the fermionic response over such backgrounds has also extensively been investigated in holographic approach since the seminal work in \cite{Faulkner:2009wj,Faulkner:2010zz}. Taking the fermions as a probe over such a background, the holographic method has been proven particularly powerful in analyzing non-Fermi liquid behavior and fermionic spectral functions that can be directly compared with experimental measurements from Angle-Resolved Photoemission Spectroscopy (ARPES) and Scanning Tunneling Microscopy (STM) \cite{Liu:2009dm}.

The early holographic study of a fermionic system is based on homogeneous backgrounds with preserved translational symmetry \cite{Lee:2008xf,Iqbal:2008by,Liu:2009dm,Faulkner:2009wj,Faulkner:2009am,Cubrovic:2009ye,Faulkner:2010zz,Wu:2011cy,Li:2011sh,Fang:2012pw,Li:2012uua}. %,Liu:2012tr}.
Recent work in this setup can also be found in \cite{Chagnet:2022ykl,Lu:2024qxj,Lu:2025zxq}. Since the lattice is an essential structure for any practical material to generate momentum relaxation and exhibit energy band structure, it is quite natural to develop the holographic fermionic system over such a lattice background that explicitly breaks the translational symmetry in spatial directions \cite{Liu:2012tr,Ling:2013aya,Ling:2014bda,Bagrov:2016cnr,Cremonini:2018xgj,Cremonini:2019fzz,Balm:2019dxk,Jeong:2019zab,Iliasov:2019pav,Mukhopadhyay:2020tky,Hercek:2022wyu}. In the presence of the lattice, two prominent features of the Fermi surface are observed in the holographic models. One is the deformation of the Fermi surface, which could be deformed from a circle to an ellipse, and the other one is the emergence of the band gap on the boundary of Brillouin Zones. Now, with the development of background geometry, we are urged to explore the fermionic response over such a background with abundant phase structure. Unfortunately, up to date the Fermionic behavior over spatially modulated backgrounds with CDW is rarely studied in literature, and the relevant work can be found only in \cite{Cremonini:2018xgj,Mukhopadhyay:2020tky}. To keep in balance, in this paper we intend to push the study on holographic fermions forward by investigating the spectral function of fermions over the background in the presence of both lattice and CDW. In contrast to the sequence adopted in \cite{Mukhopadhyay:2020tky}, where one generates the CDW at first and then turn on the lattice, we will consider the formation of CDW over a lattice background and investigate the influence of CDW on the Fermi surface. This sequence is more closely aligned with the actual situation happened in real materials since the lattice is an intrinsic structure of the matter. Furthermore, the coexistence of the lattice and CDW leads to commensurate states and incommensurate states \cite{Bednorz:1986tc,Andrade:2017leb,Andrade:2017ghg,Ling:2023ncu}, thus we intend to know how the Fermi surface changes with different commensurate ratios. In addition, the role of doping in a combined lattice-CDW system remains underexplored holographically in all previous literature. Understanding how the Fermi surface changes as the function of doping parameter in such combined systems is crucial for one to connect holographic models with experimental observations. Particularly, in this work we will focus on the following issues:
\begin{itemize}
\item The deformation of Fermi surface: How does the presence of CDW over a lattice background affect the spectral function and the structure of the Fermi surface within the first Brillouin zone?
\item The formation of the band gap: Usually the energy band gaps are developed at the boundary as the Fermi surface crosses the Brillouin zone. How does the interplay between explicit (lattice) and spontaneous (CDW) symmetry breaking lead to change of the band gaps?
\item The doping effects: How does the variation of the doping parameter influence the balance between different symmetry-breaking effects and the resulting fermionic spectral functions?
\item The dependence of the lattice amplitude: What is the role of ionic lattice strength in modifying CDW-induced effects on fermionic excitations?  In particular, as the lattice amplitude becomes large enough, the lock-in effect is expected in commensurate states. Is there any feature which could be reflected in the structure of the Fermi surface?
\end{itemize}

Based on the above issues we perform the numerical analysis on the spectral function of the fermions over the background with ionic lattice and CDW, and the main results can be summarized as follows. Firstly, we reveal that the interplay between ionic lattices and CDW enhances the amplitude of the spectral function but leads to band gap reduction through charge screening effects. Secondly, the CDW-induced charge redistribution partially compensates the ionic lattice potential, resulting in systematically smaller band gaps at Brillouin zone boundaries compared to the pure lattice case. We also find that the Fermi momentum is enhanced compared to the system with only one form of symmetry breaking, demonstrating the complex non-additive nature of combined symmetry-breaking mechanisms. Thirdly, we demonstrate that the radius of the Fermi surface expands with the increase of the doping parameter and finally cross the first Brillouin zone, developing rich band structures with multiple gaps and Fermi pockets. The above results are expected to offer direct connections to experimental observations in materials where both lattice effects and charge ordering play crucial roles, particularly in the context of high-temperature superconductors and related quantum materials.

The paper is organized as follows. In Section \ref{sec:setup}, we present our holographic setup and describe the background solutions with both ionic lattices and CDWs. Section \ref{sec:numerical} we consider the Fermions as probes over such a background and derive the spectral function from the Dirac equations. Our numerical results for the spectral function and the Fermi surface are presented in Section \ref{sec:results}. Finally, we conclude the paper with a discussion on their implications in Section \ref{sec:dis}.

\section{The holographic setup}\label{sec:setup}
We start from a gravity model with two gauge fields plus a dilaton field in four dimensions \cite{Donos:2013gda,Ling:2020qdd,Li:2024ybq}
\begin{equation}\label{eq:1}
\begin{aligned}
S= & \frac{1}{2 \kappa^2} \int d^4 x \sqrt{-g}\left[R-\frac{1}{2}(\nabla \Phi)^2-V(\Phi)-\frac{1}{4} Z_A(\Phi) F^2-\frac{1}{4} G^2\right],
\end{aligned}
\end{equation}
where $F=d A, G=d B$, $Z_A(\Phi)=1-\frac{\beta}{2} L^2 \Phi^2$, and $V(\Phi)=-\frac{1}{L^2}+\frac{1}{2} m_s^2 \Phi^2$. The first gauge field $A$ is introduced to have the notion of doping, while the second gauge field $B$ will be treated as the electromagnetic field. The dilaton field $\Phi$ plays a crucial role in inducing the instability of the background and will be treated as the order parameter of the CDW. Throughout this paper, we will set the AdS radius $l^2=6L^2=\frac{1}{4}$, the Newton's constant $2\kappa^2=1$, and the mass parameter $m_s^2=-\frac{2}{l^2}=-8$. Without loss of generality, we will also fix the adjustable parameter $\beta$ to be $\beta=-129$, which is strong enough to induce the instability of the lattice background to generate CDW.

We adopt the following ansatz for the black brane solutions with ionic lattices \cite{Ling:2013nxa,Ling:2013aya,Horowitz:2012ky,Horowitz:2012gs,Horowitz:2013jaa},
\begin{equation}
\begin{aligned}\label{eq:2}
& d s^2=\frac{1}{z^2}\left[-(1-z) p(z) Q_{t t} d t^2+\frac{Q_{z z} d z^2}{(1-z) p(z)}+Q_{x x}\left(d x+z^2 Q_{x z} d z\right)^2+Q_{y y} d y^2\right], \\
& A_t=\mu_1(1-z) a, \quad B_t=\mu_2(x)(1-z) b, \quad \Phi=z \phi, 
\end{aligned}
\end{equation}
with
\[p(z)=4\left(1+z+z^2-\frac{\left(\mu_1^2+\mu_2^2\right) z^3}{16}\right).\]
In this ansatz eight variables $\{Q_{tt}, Q_{zz}, Q_{xx}, Q_{xz}, Q_{yy}, a, b, \phi\}$ are functions of $x$ and $z$, and need to be solved from the equations of motion subject to the regular condition at the horizon $z=1$ as well as the boundary condition at $z=0$, where the asymptotical AdS is assumed. In particular, we require $Q_{tt}(x,1)=Q_{zz}(x,1)$ at the horizon such that the Hawking temperature is always given by $T=p(1)/(4 \pi)$ even in the presence of the lattice and CDW. 

The background with ionic lattices may be obtained by setting the chemical potential as $\mu_2(x)=\mu_1[X+\lambda \cos(k x)]$, which mimics how electrons experience potential from an array of ions. The doping parameter is defined as $\text{X}=\mu_2/\mu_1$ \cite{Kiritsis:2015hoa}. We remark that one advantage of this setup for $\mu_2$ is that one is allowed to set the doping parameter $X$ and the strength of the ionic lattice $\lambda$ independently. Especially, even for $X=0$, the nontrivial lattice background with $\lambda\ne 0$ may be constructed, vice versa. Furthermore, throughout this paper we adopt the chemical potential of gauge field $A$, namely $\mu_1$, as the unit. Therefore, as the background with ionic lattices, each black brane solution is characterized by a family of four dimensionless parameters: the doping parameter $X$, the temperature $T/\mu_1$, the lattice amplitude $\lambda $ as well as the wave-vector $k/\mu_1$. 

Given the lattice background with $(X,\lambda, T/\mu_1,k/\mu_1)$, the instability may be induced by the coupling term $Z_A(\Phi) F^2$ below some critical temperature, such that the system  exhibits spontaneous stripes of CDW which is characterized by a new wave-vector $p/\mu_1$, as discussed in \cite{Ling:2023ncu, Li:2024ybq}. When the wave-vector of CDW $p/\mu_1$ has a rational (irrational) ratio with the wave-vector $k/\mu_1$ of the ionic lattice, the commensurate (incommensurate) state may be constructed, which provides the basis for us to investigate the nature of Mott insulator in holographic approach. In the asymptotically AdS spacetime, to guarantee that the CDW is spontaneously generated we need to turn off the source term in $\phi$ such that near the boundary $z=0$, $B_t(x,z)$ and $\phi(x,z)$ behave as,
\begin{equation}
\begin{aligned}
B_t(x,z) &\approx \mu_{2}(x)-\rho_B(x) z + \cdots , \\
   \phi(x,z) &\approx   \phi_{2}(x) z + \cdots .
\end{aligned}
\end{equation}
In \cite{Ling:2023ncu} the effect of commensurate lock-in between the CDW and the ionic lattice is observed when the lattice amplitude is large enough \footnote{The setup in our current paper is a little bit different from the setup in \cite{Ling:2023ncu}, where the coupling between two gauge fields are taken into account.}. While in \cite{Li:2024ybq}, it is observed that two distinct types of CDW with the same commensurate rate $p/k$ may be constructed below the critical temperature, where $p$ corresponds to the CDW wave vector.
For the commensurate case $p/k=1{:}1$, the Fourier expansions of $\phi_{2}(x)$ and $\rho_{B}(x)$ are
\begin{equation}
\begin{aligned}
\phi_{2}(x) &= \phi_{2}^{(0)}+\phi_{2}^{(1)}\cos(px)+\phi_{2}^{(2)}\cos(2px)+\phi_{2}^{(3)}\cos(3px)+\cdots,\\
\rho_{B}(x) &= \rho_{B}^{(0)}+\rho_{B}^{(1)}\cos(px)+\rho_{B}^{(2)}\cos(2px)+\rho_{B}^{(3)}\cos(3px)+\cdots.
\end{aligned}
\end{equation}
For the $p/k=1{:}2$ case, the expansions take the form
\begin{equation}
\begin{aligned}
\phi_{2}(x) &=\phi_{2}^{(1)}\cos(px)+\phi_{2}^{(3)}\cos(3px)+\phi_{2}^{(5)}\cos(5px)+\cdots,\\
\rho_{B}(x) &= \rho_{B}^{(0)}+\rho_{B}^{(2)}\cos(2px)+\rho_{B}^{(4)}\cos(4px)+\rho_{B}^{(6)}\cos(6px)+\cdots.
\end{aligned}
\end{equation}
Since the setup in our current paper is identical to that prior to the superconductivity condensation in \cite{Li:2024ybq}, we refer to \cite{Li:2024ybq} for more details on the construction of the lattice background with CDW. The phase diagram on $X-T$ plane with various values of the lattice amplitude is also available in \cite{Li:2024ybq}.

\section{Fermions }\label{sec:numerical}
In order to analyze how fermionic operators behave in the dual field theory, we need to examine their spectral functions. To accomplish this, we will consider generic fermions in the gravitational theory and analyze their equations of motion through the Dirac equation formalism. Throughout this analysis, we work in the probe limit, neglecting the gravitational backreaction of the fermions on the background geometry. The bulk action for the Dirac field with mass $m$ and charge $q$ is given by \cite{Ling:2013aya}
\begin{equation}\label{eq:3}
S_D=i \int d^4 x \sqrt{-g}\bar{\zeta}\left(\Gamma^a \mathcal{D}_a-m\right) \zeta,
\end{equation}
in the static background geometry
\begin{equation}
  \begin{aligned}\label{eq:6}
  d s^2 & =-g_{t t}(x, z) d t^2+g_{z z}(x, z) d z^2+g_{x x}(x, z) d x^2+g_{y y}(x, z) d y^2+2 g_{z z}(x, z) d x d z \\
  B & =B_t(x, z) d t ,
  \end{aligned}
  \end{equation}
where the covariant derivative is $\mathcal{D}_a=\partial_a+\frac{1}{4}\left(\omega_{\mu \nu}\right)_a \Gamma^{\mu \nu}-i q B_a$ with $\Gamma^{\mu \nu}=\frac{1}{2}\left[\Gamma^\mu, \Gamma^\nu\right]$. Here the gamma matrices are given by $\Gamma^a=\left(e_\mu\right)^a \Gamma^\mu$ with $\left(e_\mu\right)^a$ being a set of orthogonal normal vector bases and the spin connection 1-forms $\left(\omega_{\mu \nu}\right)_a=\left(e_\mu\right)_b \nabla_a\left(e_\nu\right)^b $. We also remark that since we treat $B$ as the electromagnetic field, the possible coupling between the first gauge field $A$ with the fermion is ignored. In addition, we will not consider the possible couplings between the scalar field $\phi$ and the fermions in this paper for simplicity. For the background geometry in Eq.(\ref{eq:2}), where the horizon is at $z=1$ and the AdS boundary is at $z=0$, we have flexibility in selecting the vielbein and gamma matrices. In our analysis, we adopt the following choice of orthogonal normal vector bases
\begin{equation}
  \begin{aligned}\label{eq:7}
  & \left(e_0\right)^a=\frac{1}{\sqrt{g_{t t}}}\left(\frac{\partial}{\partial t}\right)^a, \quad\left(e_1\right)^a=\frac{1}{\sqrt{g_{x x}}}\left(\frac{\partial}{\partial x}\right)^a, \quad\left(e_2\right)^a=\frac{1}{\sqrt{g_{y y}}}\left(\frac{\partial}{\partial y}\right)^a, \\
  & \left(e_3\right)^a=-\sqrt{\frac{g_{x x}}{g_{x x} g_{z z}-g_{x z}^2}}\left(\frac{\partial}{\partial z}\right)^a+\frac{g_{x z}}{\left.\sqrt{g_{x x}\left(g_{x x} g_{z z}-g_{x z}^2\right.}\right)}\left(\frac{\partial}{\partial x}\right)^a,
  \end{aligned}
\end{equation}
and the non-vanishing components of the spin connection can be calculated as follows:
\begin{equation}
  \begin{aligned}\label{eq:8}
  & \left(\omega_{01}\right)_a=-\left(\omega_{10}\right)_a=-\frac{\partial_x g_{t t}}{2 \sqrt{g_{t t} g_{x x}}}(d t)_a, \\
  & \left(\omega_{03}\right)_a=-\left(\omega_{30}\right)_a=\frac{g_{x x} \partial_z g_{t t}-g_{x z} \partial_x g_{t t}}{2 \sqrt{g_{t t} g_{x x}\left(g_{x x} g_{z z}-g_{x z}^2\right)}}(d t)_a, \\
  & \left(\omega_{12}\right)_a=-\left(\omega_{21}\right)_a=-\frac{\partial_x g_{y y}}{2 \sqrt{g_{x x} g_{y y}}}(d y)_a, \\
  & \left(\omega_{13}\right)_a=-\left(\omega_{31}\right)_a=\left(-\frac{\partial_z g_{x x}}{2 \sqrt{g_{x x} g_{z z}-g_{x z}^2}}+\frac{2 g_{x x} \partial_x g_{x z}-g_{x z} \partial_x g_{x x}}{2 g_{x x} \sqrt{g_{x x} g_{z z}-g_{x z}^2}}\right)(d x)_a \\
  & +\frac{g_{x x} \partial_{x } g_{z z}-g_{z z} \partial_z g_{x x}}{2 g_{x x} \sqrt{g_{x x} g_{z z}-g_{x z}^2}}(d z)_a, \\
  & \left(\omega_{23}\right)_a=-\left(\omega_{32}\right)_a=\frac{g_{x z} \partial_x g_{y y}-g_{x x} \partial_z g_{y y}}{2 \sqrt{g_{x x} g_{y y}\left(g_{x x} g_{z z}-g_{x z}^2\right)}}(d y)_a .
  \end{aligned}
  \end{equation}
The Dirac equation obtained from Eq.\ref{eq:3} reads
\begin{equation}
\left(\Gamma^a \mathcal{D}_a-m\right) \zeta=0.\label{eq:5}
\end{equation}
To proceed, we redefine $\zeta$ via $\zeta =(g_{tt}g_{xx}g_{yy})^{1/4}  e^{-i \omega t +ik_i x^i} F(x,z)$, then Eq.\ref{eq:5} can be expressed as
\begin{equation}
\begin{aligned}\label{eq:9}
\Delta_0 \Gamma^0 F-\Delta_1 \Gamma^1 F-\Delta_2 \Gamma^2 F+\Delta_3 \Gamma^3 F+m F=0,
\end{aligned}
\end{equation}
with
\begin{equation}
\begin{aligned}\label{eq:10}
& \Delta_0=: i\left(\omega+q B_t\right) \frac{1}{\sqrt{g_{t t}}}, \\
& \Delta_1=:\left(\frac{1}{\sqrt{g_{x x}}} \partial_x+\frac{i k_1}{\sqrt{g_{x x}}}-\frac{\partial_{x } g_{x x}}{4 g_{x x}^{3 / 2}}+\frac{\sqrt{g_{x x}} \partial_{x } g_{z z}}{4\left(g_{x x} g_{z z}-g_{x z}^2\right)}+\frac{g_{x z}\left(g_{x z} \partial_{x } g_{x x}-2 g_{x x} \partial_{x x} g_{x z}\right)}{4 g_{x x}^{3 / 2}\left(g_{x x} g_{z z}-g_{x z}^2\right)}\right) \\
& \Delta_2=: \frac{i k_2}{\sqrt{g_{y y}}} \\
& \Delta_3=: \frac{1}{\sqrt{g_{x x}\left(g_{x x} g_{z z}-g_{x z}^2\right)}}\left(g_{x x} \partial_z-g_{x z} \partial_x-i k_1 g_{x z}-\frac{1}{2} \partial_x g_{x z}+\frac{g_{x z}}{2 g_{x x}} \partial_x g_{x x}\right)
\end{aligned}
\end{equation}
To solve the Dirac equation, we choose a specific representation of the gamma matrices in terms of Pauli matrices $\sigma^i$:
\begin{equation}
\begin{aligned}\label{eq:11}
 \Gamma^0&=\left(\begin{array}{cc}
i \sigma^1 & 0 \\
0 & i \sigma^1
\end{array}\right), 
\quad \Gamma^1=\left(\begin{array}{cc}
-\sigma^2 & 0 \\
0 & \sigma^2
\end{array}\right),  \\
\Gamma^2&=\left(\begin{array}{cc}
0 & \sigma^2 \\
\sigma^2 & 0
\end{array}\right),
\quad  \Gamma^3=\left(\begin{array}{cc}
-\sigma^3 & 0 \\
0 & -\sigma^3
\end{array}\right). 
\end{aligned}
\end{equation}
Furthermore, we decompose the spinor $F$ into two components $F=\left(F_1, F_2\right)^T$. With this choice of gamma matrices and spinor decomposition, the Dirac equation in Eq.(\ref{eq:9}) takes the form
\begin{equation}
\begin{aligned}\label{eq:12}
\Delta_0 \sigma^2 \otimes\binom{F_1}{F_2} \pm i \Delta_1 \sigma^1 \otimes\binom{F_1}{F_2}-i \Delta_2 \sigma^1 \otimes\binom{F_2}{F_1}+\Delta_3\binom{F_1}{F_2}-m \sigma^3 \otimes\binom{F_1}{F_2}=0.
\end{aligned}
\end{equation}
For convenience, we further split $F_{\alpha}(\alpha=1,2)$ as
\begin{equation}
\begin{aligned}\label{eq:12b}
F_\alpha \equiv\binom{\mathcal{A}_\alpha}{\mathcal{B}_\alpha},
\end{aligned}
\end{equation}
then the Dirac equation in Eq.(\ref{eq:11}) reduces to
\begin{equation}
\begin{aligned}\label{eq:13}
& \left(\Delta_{30} \partial_z+\Delta_{31} \partial_x+\Delta_{32} \mp m\right)\binom{\mathcal{A}_1}{\mathcal{B}_1} \mp i \Delta_0\binom{\mathcal{B}_1}{\mathcal{A}_1}+i \Delta_1\binom{\mathcal{B}_1}{\mathcal{A}_1}-i \Delta_2\binom{\mathcal{B}_2}{\mathcal{A}_2}=0, \\
& \left(\Delta_{30} \partial_z+\Delta_{31} \partial_x+\Delta_{32}\mp m\right)\binom{\mathcal{A}_2}{\mathcal{B}_2} \mp i \Delta_0\binom{\mathcal{B}_2}{\mathcal{A}_2}-i \Delta_1\binom{\mathcal{B}_2}{\mathcal{A}_2}-i \Delta_2\binom{\mathcal{B}_1}{\mathcal{A}_1}=0,
\end{aligned}
\end{equation}
where
\begin{equation}
\begin{aligned}\label{eq:14}
\Delta_{30} & =: \frac{g_{x x}}{\sqrt{g_{x x}\left(g_{x x} g_{z z}-g_{x z}^2\right)}}, \\
\Delta_{31} & =: \frac{-g_{x z}}{\sqrt{g_{x x}\left(g_{x x} g_{z z}-g_{x z}^2\right)}}, \\
\Delta_{32} & =: \frac{-i k_1 g_{x z}-\frac{1}{2} \partial_x g_{x z}+\frac{g_{x z}}{2 g_{x x}} \partial_x g_{x x}}{\sqrt{g_{x x}\left(g_{x x} g_{z z}-g_{x z}^2\right)}} .
\end{aligned}
\end{equation}

For a lattice background with period $c$ in $x$ direction, the Bloch theorem allows us to expand the Dirac spinor components as
\begin{equation}
\begin{aligned}\label{eq:15}
\binom{\mathcal{A}_\alpha(x, z)}{\mathcal{B}_\alpha(x, z)}=\sum_{n=0, \pm 1, \pm 2, \cdots}\binom{\mathcal{A}_{\alpha, n}(z)}{\mathcal{B}_{\alpha, n}(z)} e^{i n K x},
\end{aligned}
\end{equation}
with $K=\frac{2 \pi}{c}$.

Substituting the background geometry Eq.\ref{eq:2} into Eq. \ref{eq:13}, near the horizon $z=1$, we have 
\begin{equation}
\begin{aligned}\label{eq:16}
\partial_z\binom{\mathcal{A}_{\alpha, n}}{\mathcal{B}_{\alpha, n}} \pm \frac{\omega}{4 \pi T} \frac{1}{1-z}\binom{\mathcal{B}_{\alpha, n}}{\mathcal{A}_{\alpha, n}}=0,
\end{aligned}
\end{equation}
thus the solution must satisfy the infalling boundary conditions required for computing the retarded Green function:
\begin{equation}
\begin{aligned}\label{eq:17}
\binom{\mathcal{A}_{\alpha, n}}{\mathcal{B}_{\alpha, n}}=\binom{1}{-i}(1-z)^{-\frac{i \omega}{4 \pi T}}.
\end{aligned}
\end{equation}
for each mode $(\alpha,n)$ while setting other modes to zero. On the other hand, near the AdS boundary, the asymptotic expansion of Eq. (\ref{eq:13}) takes the form
\begin{equation}
\begin{aligned}\label{eq:18}
\left(z \partial_z-m \sigma^3\right) \otimes\binom{F_{1, n}}{F_{2, n}}=0 .
\end{aligned}
\end{equation}
Hence the solution can be asymptotically expanded near the AdS boundary as
\begin{equation}
\begin{aligned}\label{eq:19}
F_{\alpha, n} \approx a_{\alpha, n} z^m\binom{1}{0}+b_{\alpha, n} z^{-m}\binom{0}{1},
\end{aligned}
\end{equation}
where $a_{\alpha,n}$ and $b_{\alpha,n}$ are the source and response coefficients respectively. The retarded Green's function can then be extracted from
 \begin{equation}\label{eq:20}
a_{\alpha, n}(\beta, l)=G_{\alpha, n ; \alpha^{\prime}, n^{\prime}} b_{\alpha^{\prime}, n^{\prime}}(\beta, l),
\end{equation}
where the coefficients correspond to solutions with only the $(\beta,l)$ mode excited at the horizon.

The diagonal momentum spectral weight is defined as 
\begin{equation}
\mathcal{G}=\sum_{n=0, \pm 1, \pm 2, \cdots} \operatorname{Tr} \operatorname{Im}\left[G_{\alpha, n ; \alpha^{\prime}, n^{\prime}}\right] .
\end{equation}
We aim to investigate the spectral function of generic fermions in the background with ionic lattice which may undergo a phase transition with the development of charge density wave by the spontaneous breaking of the translational symmetry. For simplicity, we restrict our attention to the massless case ($m=0$). Since analytical solutions are not available, we employ numerical methods to solve the Dirac equations with various infalling boundary conditions, from which we extract the spectral density function.

\section{Results}\label{sec:results}
In what follows, we investigate the structure of Fermi surface through the analysis of the spectral function across various parameter regimes. It is important to note that while the strict definition of a Fermi surface requires zero temperature and unbroken translational symmetry, one may still meaningfully identify Fermi surface features in low-temperature lattice system. This is possible because, as demonstrated by ARPES experiments \cite{Liu:2012tr}, the lattice effects are effectively averaged out into a continuum. Specifically, the spectral function can be extracted from the imaginary part of the diagonal components of the retarded Green's function:
\begin{equation}
A(\omega, k_x = k_1 + nK, k_y = k_2) = \text{Im}(G_{1,n;1,n} + G_{2,n;2n}).
\end{equation}
Given that the ultra-cold lattice introduces only minimal smearing of the Fermi surface, we can precisely locate the Fermi surface by identifying peaks in $A(\omega, k_x, k_y)$ at infinitesimal frequencies $\omega$ in momentum space. This methodology for Fermi surface identification aligns with previous approaches \cite{Benini:2010qc,Kanigel}.

While the existence of Fermi surfaces in holographic fermionic liquids is well-established (see \cite{Faulkner:2011tm,Iqbal:2011ae} for comprehensive reviews), in this section we will study the structure of Fermi surface and the band gap when crossing different Brillouin zones.

\subsection{The spectral function and Fermi surface in the first Brillouin zone}

In this subsection, we study the spectral function and the Fermi surface for various parameters within the first Brillouin zone and explore its deformation over the background with ionic lattices and CDWs. Throughout this paper, we will fix the temperature at $T/\mu_1=0.005$, which is low enough to observe the zero-temperature behavior of Fermions. Moreover, in the presence of CDW we will focus on the commensurate states with $p/k=1:1$ in these two subsections, but leave the states with $p/k=1:2$ for discussion in subsection \ref{subsectC}.
Thus the gravitational background with CDW is now characterized by three dimensionless parameters ($X$, $\lambda$, $k/\mu_1$), while the coupled fermion is massless and solely characterized by its charge $q$.
\begin{figure} [h]
  \center{
    \includegraphics[width=0.45\textwidth]{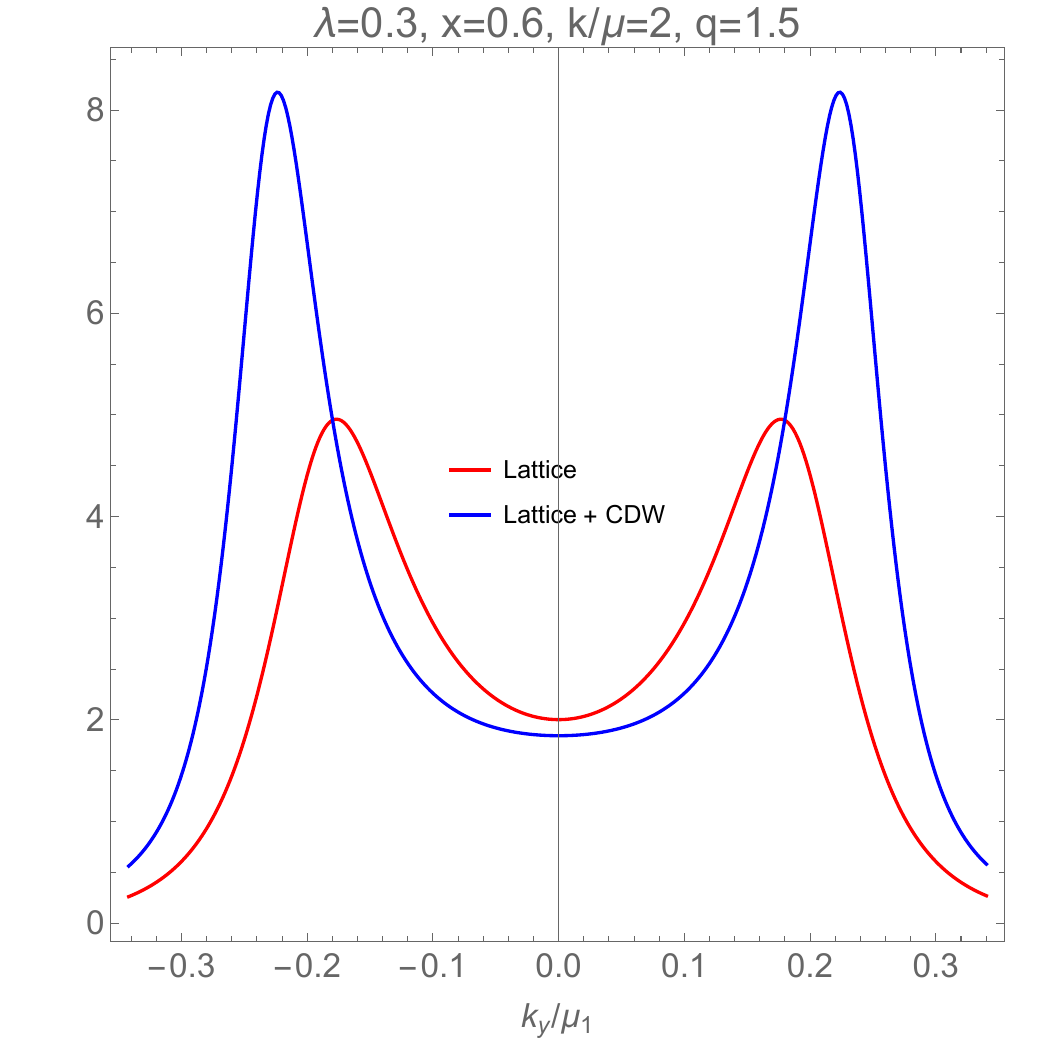}\ \hspace{0.8cm}
    \includegraphics[width=0.45\textwidth]{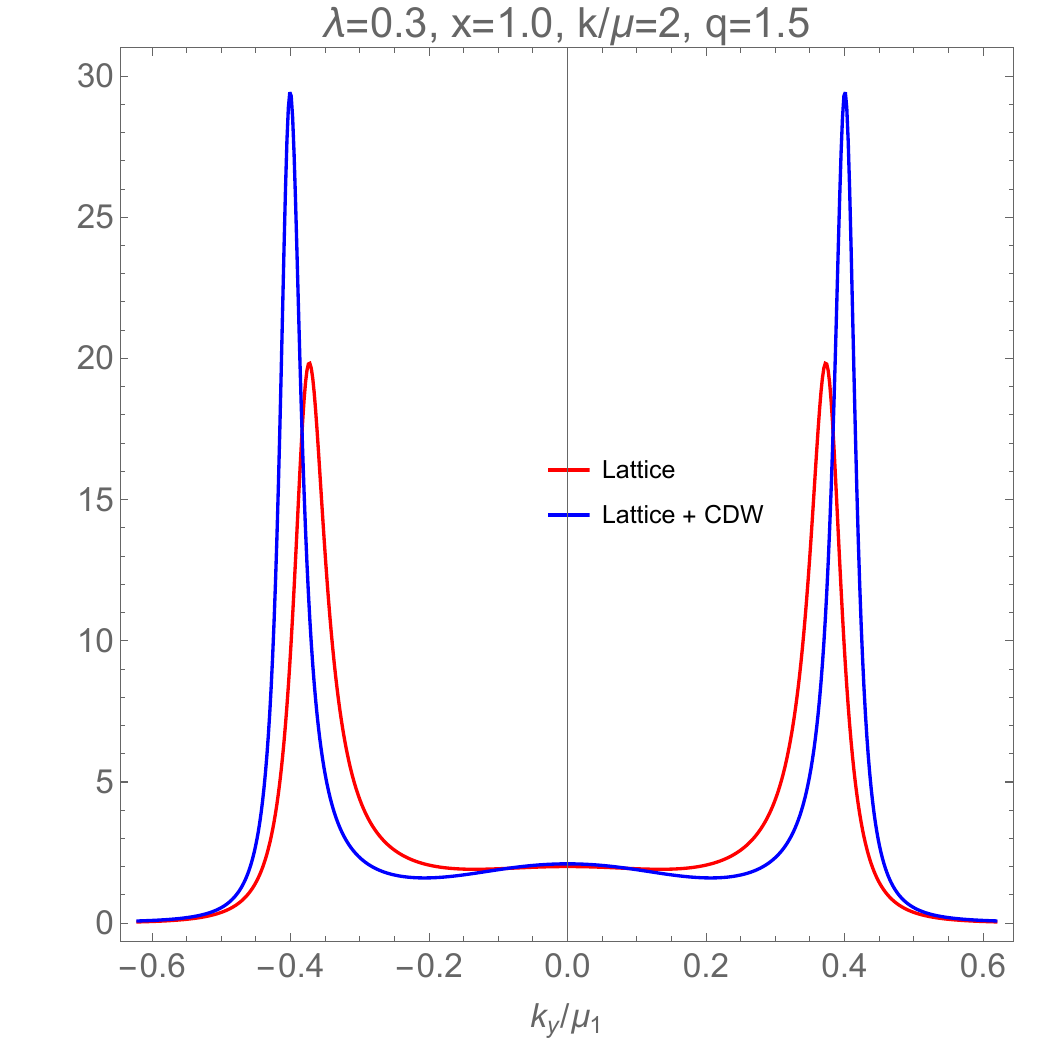}\ \hspace{0.05cm}
    \caption{\label{fig-2}   Spectral function $A(\omega\!\to\!0,\mathbf{k})$ %at $T/\mu_1=0.005$ 
    for different doping $X$, where $k_x/\mu_1=0$.}}
\end{figure}

Firstly, we demonstrate the structure of the spectral function, with a focus on the interplay between ionic lattice and CDW. As a typical example, we plot the spectral function $A(\omega,k)$ as the function of $k_y/\mu_1$ within the first Brillouin zone in Fig. \ref{fig-2} with $k_x/\mu_1=0$. Without surprise, the spectral function exhibits a mirror symmetry when $k_y\rightarrow -k_y$. The sharp peak indicates the location of the Fermi momentum. The left panel is plotted for the doping $X = 0.6$ while the right panel is for $X = 1.0$. In each panel, it is noticed that the presence of CDW makes the peak of the spectral more pronounced and the corresponding Fermi momentum becomes larger. Furthermore, comparing the spectral function in these two panels, we find that the peaks of the spectral function become sharper and more pronounced at higher doping levels. This enhancement of spectral weight indicates increased quasiparticle coherence, which can be attributed to the suppression of CDW order at higher doping levels (as evidenced by the reduced critical temperature with increasing $X$ \cite{Li:2024ybq}). The weaker CDW modulation leads to reduced Umklapp scattering, thereby enhancing the quasiparticle lifetime and resulting in sharper spectral features. Here, Umklapp scattering refers to a scattering process where the momentum transfer equals the reciprocal lattice vector $K = 2\pi/L$, causing electrons to scatter between different Brillouin zones. Within the first Brillouin zone, although the Fermi surface does not cross the zone boundary, the coupling to Umklapp modes ($k \pm K$) contributes to the quasiparticle self-energy. This manifests primarily as a correction to normal scattering processes, leading to spectral broadening but preserving the continuity of the Fermi surface. Therefore, the observed sharpening of the spectral peaks at higher doping can be directly identified with the suppression of this scattering channel, signaling a recovery of the quasiparticle coherence.

The systematic enhancement of spectral function with increasing doping parameter $X$ observed in Fig. \ref{fig-2} reveals a fundamental competition between kinetic energy and interaction-driven ordering in our system. This behavior can be understood through two interconnected mechanisms:
\begin{itemize}
    \item CDW Order Parameter Suppression: From the phase diagram established in \cite{Li:2024ybq}, the critical temperature of forming CDW decreases systematically with the increase of doping parameter $X$. At our working temperature, which is set sufficiently low to capture near-ground-state behavior, the system exhibits distinct behaviors across the doping range. At lower doping levels, the system operates deep within the CDW phase, with the working temperature being substantially below the critical temperature. In contrast, at higher doping levels, the system approaches much closer to the phase boundary in the phase diagram, with the working temperature representing a larger fraction of the critical temperature. This proximity to the critical point directly translates to weaker CDW modulation strength, which in turn reduces the Umklapp scattering rate experienced by fermions. The sharper spectral peaks at higher $X$ reflect enhanced quasiparticle coherence resulting from this suppression.
    \item Enhanced Screening by Mobile Carriers: As doping increases, more mobile charge carriers become available to screen the periodic potentials in the system. The boundary condition $\mu_2(x)=\mu_1[X+\lambda \cos(k x)]$ implies that the average charge density grows proportionally with $X$, i.e., $\langle \rho \rangle \propto X$. These additional carriers act collectively to ``smooth out'' both the ionic lattice potential and the CDW-induced charge modulation, similar to how mobile electrons in a metal screen external electric fields. This screening effect manifests through two mechanisms: (i) the effective screening length decreases as $\lambda_{\text{screen}} \sim 1/\sqrt{X}$, allowing the system to more efficiently neutralize potential variations at higher doping; (ii) while the absolute magnitude of CDW-induced charge redistribution $\Delta\rho(x)$ may remain finite, its relative importance diminishes—the ratio $\Delta\rho(x)/\langle\rho\rangle$ decreases with increasing $X$. Consequently, the CDW modulation becomes increasingly ``diluted'' by abundant mobile carriers, weakening its influence on fermionic excitations and contributing to enhanced spectral coherence.
\end{itemize}

These two mechanisms work synergistically: CDW order parameter suppression reduces the intrinsic strength of charge modulation (through proximity to $T_c$), while enhanced screening diminishes the effective impact of this modulation on fermionic excitations (through charge redistribution). Together, they drive the system from a CDW-dominated strongly correlated regime at low $X$ to a weakly correlated itinerant regime at high $X$, where kinetic energy becomes increasingly competitive with interaction-driven ordering. This behavior mirrors the doping evolution observed in cuprate superconductors, where the pseudogap regime (analogous to our low-$X$ region) gives way to Fermi-liquid-like behavior upon increased doping, providing a holographic realization of this experimental phenomenology \cite{boebinger1996insulator,Vojta01112009}.

\begin{figure} [h]
  \center{
    \includegraphics[width=0.45\textwidth]{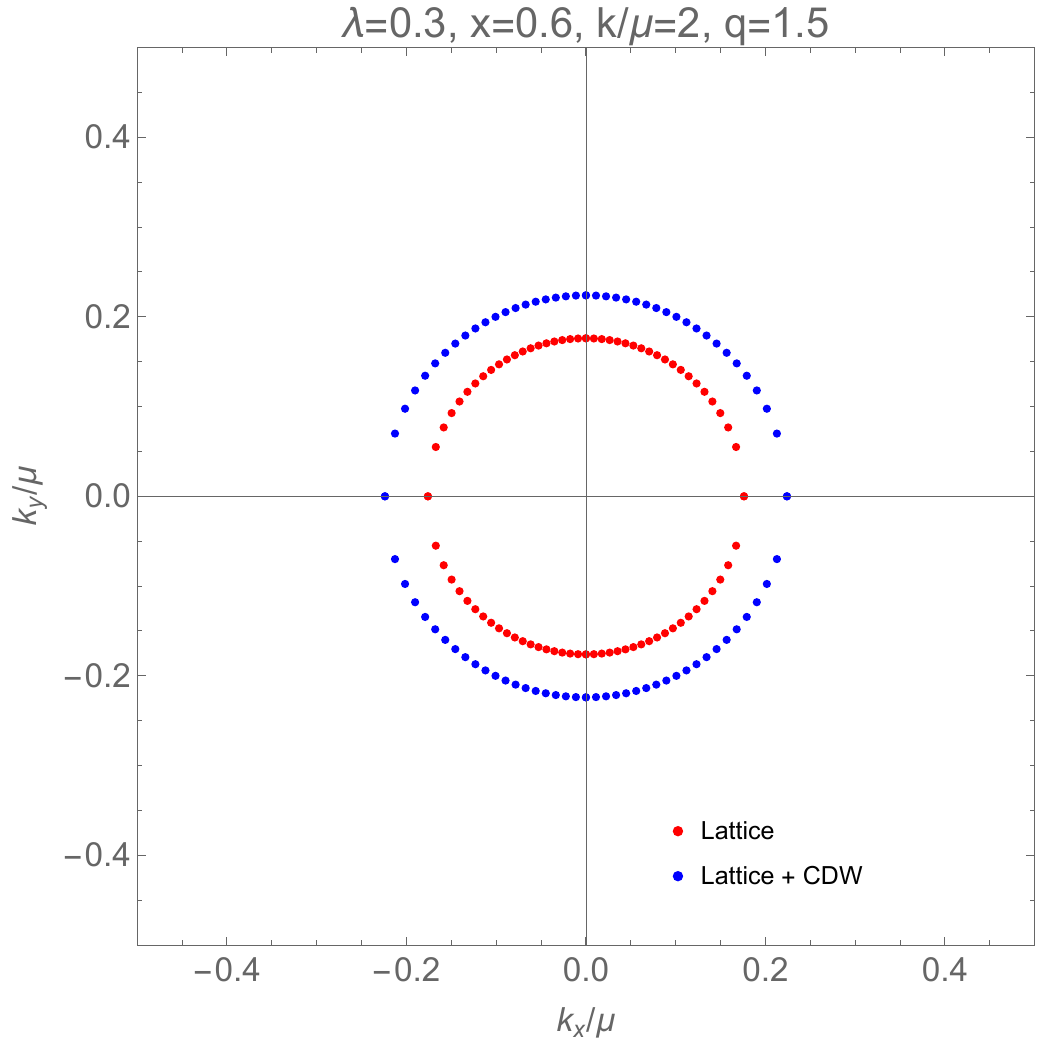}\ \hspace{0.8cm}
    \includegraphics[width=0.45\textwidth]{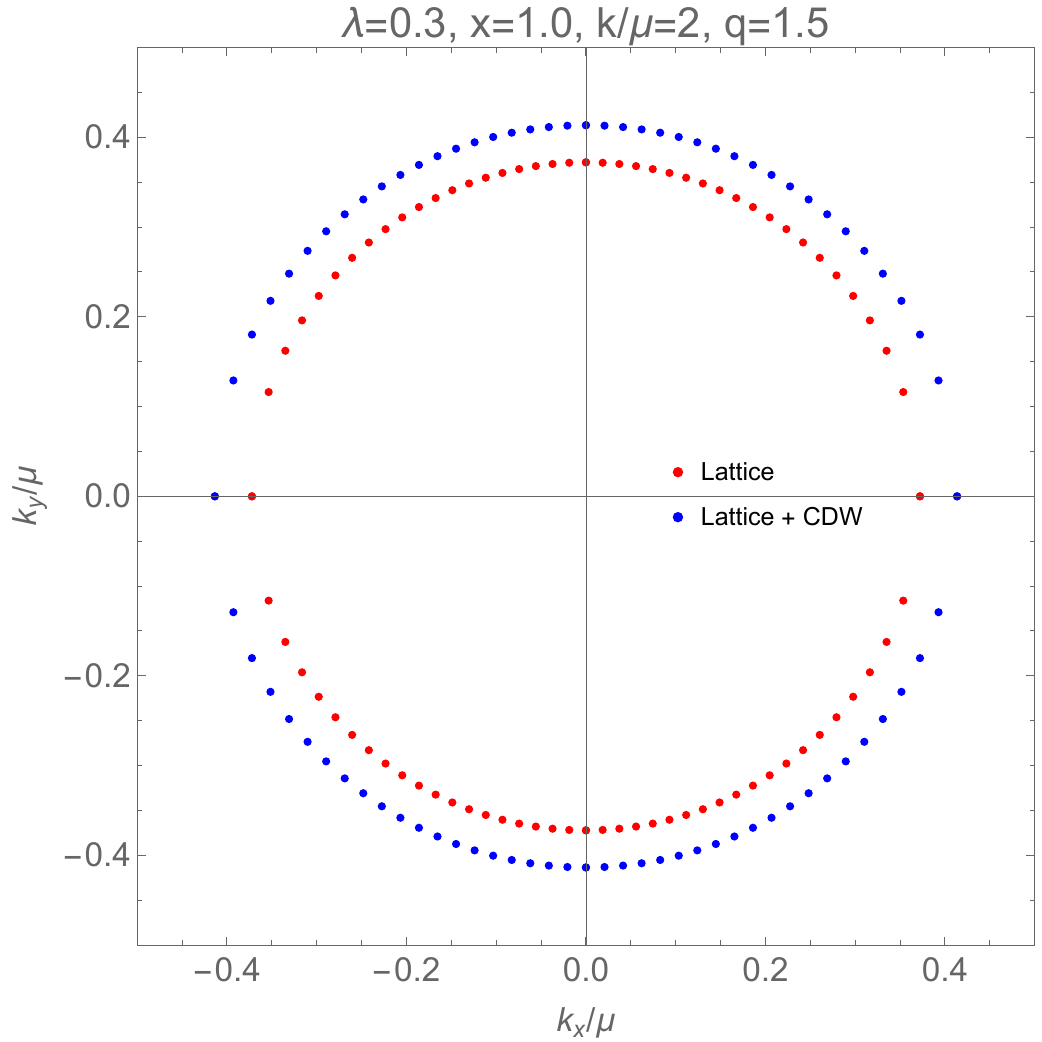}\ \hspace{0.05cm}
    \caption{\label{fig-3} The Fermi surface for $X=0.6$ (left) and $X=1.0$ (right).
    }}
\end{figure}

Next, we turn to the structure of the Fermi surface. At strictly zero temperature, the Fermi surface manifests as poles in the spectral function. While in the present case of very low but non-zero temperature, we identify the location of the Fermi surface as the pronounced peaks in the spectral function, which is captured by the imaginary part of the diagonal components of the retarded Green function. The ultra-cold lattice introduces only minimal smearing, allowing clear identification of these Fermi surface features in momentum space. Identifying the peak positions $k_y/\mu_1$ of the spectral function as the Fermi momentum and scanning the peaks with different values of $k_x/\mu_1$, one can construct the complete Fermi surface in momentum space. It is known that in a homogeneous system with spatially rotational symmetry, the Fermi surface typically forms a perfect circle in momentum space. However, when translational symmetry is broken along one direction (as in our model with ionic lattice and/or CDW), the rotational symmetry is also broken. Consequently, the Fermi surface is deformed from its conventional circular shape to an elliptical geometry, which has been elegantly demonstrated in \cite{Ling:2013aya}. This elliptical shape is remarkably universal and robust across different parameter regimes. 

Fig. \ref{fig-3} provides a clear visualization of how the Fermi surface changes with the doping parameter $X$ within the first Brillouin zone. Two plots demonstrate the Fermi surface for $X = 0.6$, and $X=1.0$ respectively, with all other parameters fixed as $(\lambda = 0.3, q = 1.5, k/\mu = 2)$. In each panel, two distinct Fermi surfaces are plotted: the inner contour (labeled ``Lattice'') represents the Fermi surface for a system with ionic lattice only, while the outer contour (labeled ``Lattice and CDW'') corresponds to the Fermi surface for a system with both ionic lattice + CDW with the same parameter values. For clarity, we summarize the rules observed in this figure as the following list.

\begin{itemize}
    \item First of all, we notice that as $X$ increases from 0.6 to 1.0, both Fermi surfaces expand evidently, indicating an increase in the number of charge carriers with higher doping levels.
    \item The Fermi surface in Fig. \ref{fig-3} exhibits slight elliptical deformation due to broken translational symmetry along the $x$-direction. However, for the parameters studied ($\lambda=0.3, k/\mu_1=2$), the deviation from circular symmetry is modest. The tendency for anisotropic dispersion reflects enhanced effective mass along the modulation direction, though this effect becomes more pronounced at lower doping levels or larger lattice amplitudes where symmetry-breaking effects are stronger.
    \item The comparison between the inner and outer contours in Fig. \ref{fig-3} reveals that the presence of CDW consistently increases the Fermi surface size compared to the lattice-only case. While both exhibit elliptical shapes due to broken translational symmetry along the $x$-direction, the CDW introduces additional charge redistribution that expands the Fermi momentum. This enhancement is particularly pronounced at lower $X$ values ($X = 0.6$), suggesting that CDW effects are stronger at lower doping levels. As $X$ increases to $1.0$, the difference between the lattice-only and lattice+CDW Fermi surfaces becomes less pronounced, consistent with the expectation that higher doping suppresses the CDW phase, as shown in the phase diagram where the critical temperature of CDW decreases with the increase of $X$ \cite{Li:2024ybq}. Despite this suppression, the elliptical shape persists across all doping levels, highlighting its robustness as a feature of systems with broken translational symmetry.
\end{itemize}
Finally, we remark that our system—featuring coexisting ionic lattice and CDW from the outset—exhibits distinct Fermi surface characteristics compared to previous holographic studies. The combined explicit (lattice) and spontaneous (CDW) symmetry breaking leads to continuous elliptical Fermi surface contours throughout the first Brillouin zone. This geometry remains robust across different doping levels, with the elliptical deformation becoming less pronounced at higher doping where CDW effects are suppressed. Additionally, unlike Ref.\cite{Cremonini:2018xgj} where strong lattice effects lead to spectral weight suppression and Fermi arc formation, our moderate lattice amplitude ($\lambda=0.3$) preserves closed Fermi surfaces with well-defined peaks in the spectral function.
\begin{figure} [h]
  \center{
    \includegraphics[width=0.6\textwidth]{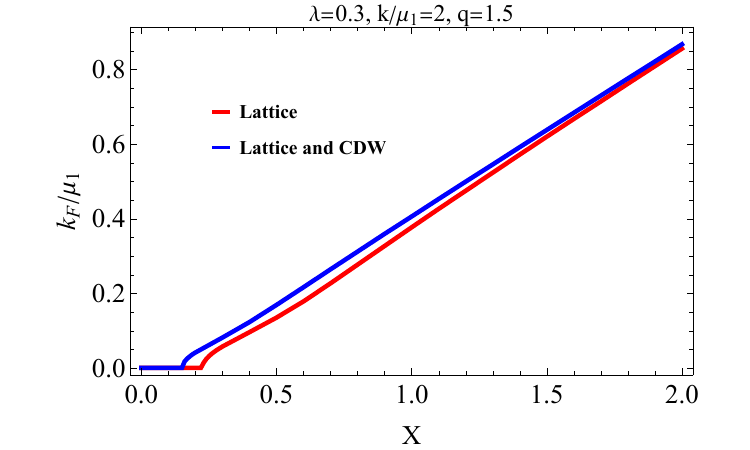}
    \caption{\label{fig-3-2}
    The Fermi momentum $k_F/\mu_1$ versus the doping parameter $X$ when other parameters are fixed.}}
\end{figure}

To quantitatively demonstrate these trends, Fig. \ref{fig-3-2} plots the Fermi momentum $k_F/\mu_1$ versus the doping parameter $X$. This figure reveals several crucial aspects of the crossover from strongly correlated to weakly correlated regimes in our system.

First, a finite threshold in $X$ must be exceeded before a well-defined Fermi surface with non-zero $k_F$ emerges, indicating that a minimum carrier density is required for Fermi surface formation. This threshold behavior reflects the competition between interaction-driven localization at low doping and the emergence of itinerant quasiparticles at higher carrier densities.

Second, Fig. \ref{fig-3-2} demonstrates an intriguing approximately linear scaling $k_F/\mu_1 \propto X$, which deviates significantly from the $k_F \propto \sqrt{n}$ relation expected for weakly interacting fermions in (2+1) dimensions. This deviation constitutes a hallmark signature of strong coupling effects that renormalize the effective dimensionality governing the density-momentum relation. The linear behavior suggests that the system does not simply follow free-fermion kinematics even at higher doping levels, but rather exhibits persistent strong correlation effects combined with nonlinear charge redistribution induced by the coexisting lattice and CDW.

Third, comparing the ``lattice only'' and ``lattice+CDW'' curves in Fig. \ref{fig-3-2}, we observe that CDW consistently enhances the Fermi momentum across all doping levels. However, the gap $\Delta k_F = k_F(\text{lattice+CDW}) - k_F(\text{lattice})$ diminishes systematically with increasing $X$. This convergence behavior can be understood through the enhanced screening mechanism: as $X$ increases, the screening length $\lambda_{\text{screen}} \sim 1/\sqrt{X}$ decreases, rendering the CDW modulation less effective relative to the growing background charge density. The observed trend indicates a progressive crossover from a CDW-dominated regime at low $X$ to a more itinerant regime at high $X$, where kinetic energy becomes increasingly competitive with interaction-driven ordering. This doping evolution provides a holographic analog of the pseudogap-to-Fermi-liquid crossover observed in cuprate superconductors.

In the next subsection, we turn to examine how the Fermi surface behaves when it extends beyond the first Brillouin zone, focusing on the formation of band gaps at zone boundaries and their relationship with both explicit (lattice) and spontaneous (CDW) breaking of translational symmetry.

\subsection{Band Gap Structure Beyond the First Brillouin Zone}
When the Fermi surface extends beyond the first Brillouin zone, then due to the periodic structure in momentum space the band gap is expected to emerge at the Brillouin zone boundaries. Remarkably, this has been justified in holographic fermionic liquids with broken translational symmetry \cite{Liu:2012tr,Ling:2013aya}. In this subsection we intend to examine how these band gaps form at the boundary with the change of the system parameters, with a focus on the relationship between the geometry of fermion surface and the background with both ionic lattices and CDWs.
\begin{figure} [h]
  \center{
    \includegraphics[width=0.45\textwidth]{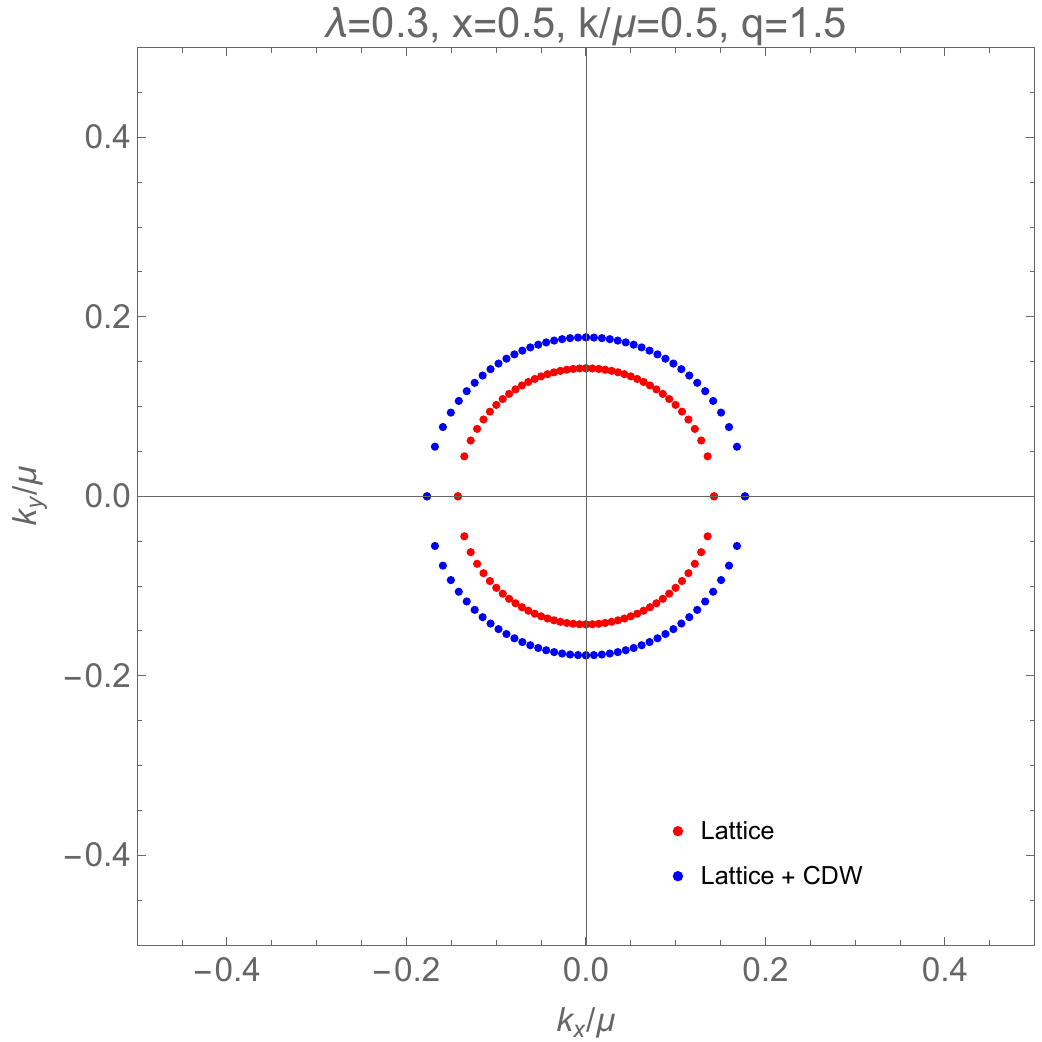}\ \hspace{0.8cm}
    \includegraphics[width=0.45\textwidth]{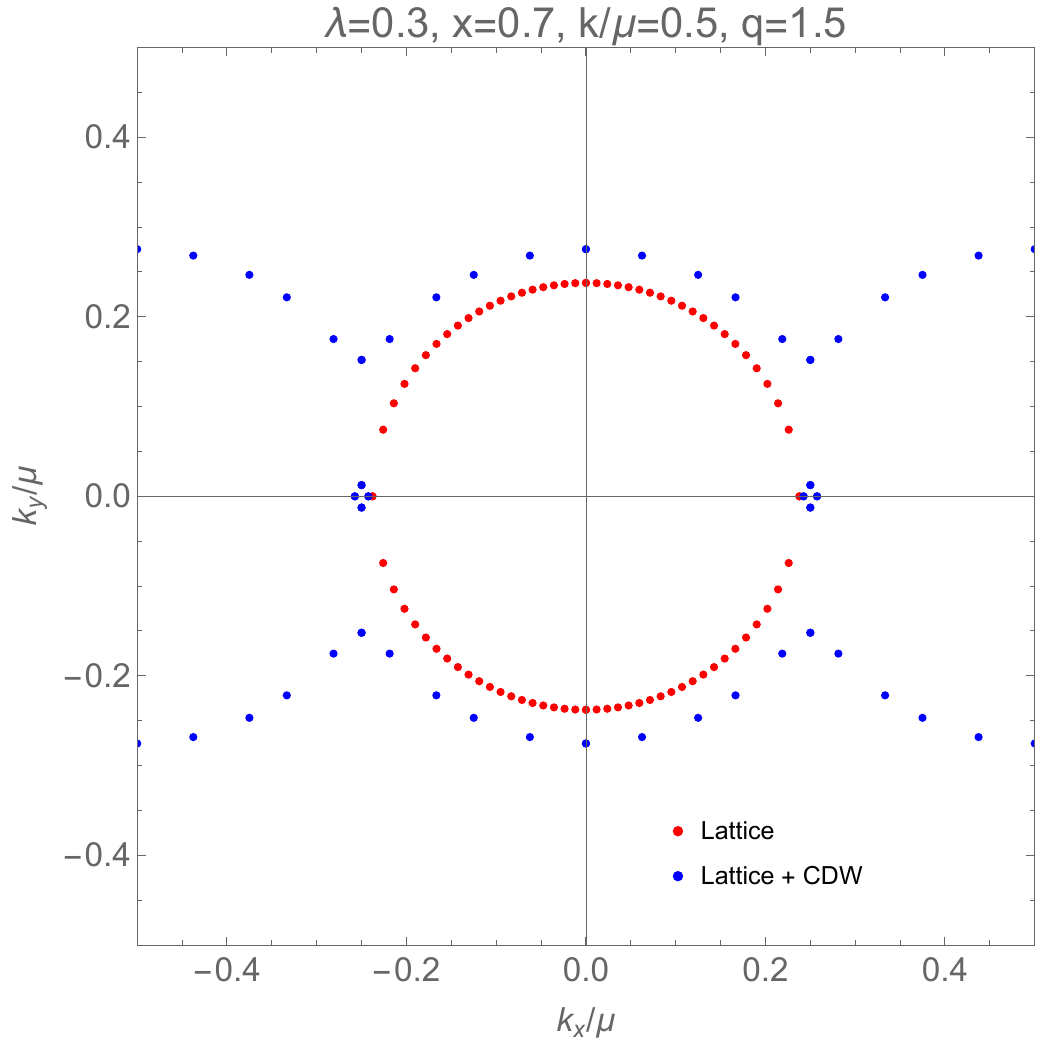}\ \hspace{0.05cm}
    \includegraphics[width=0.45\textwidth]{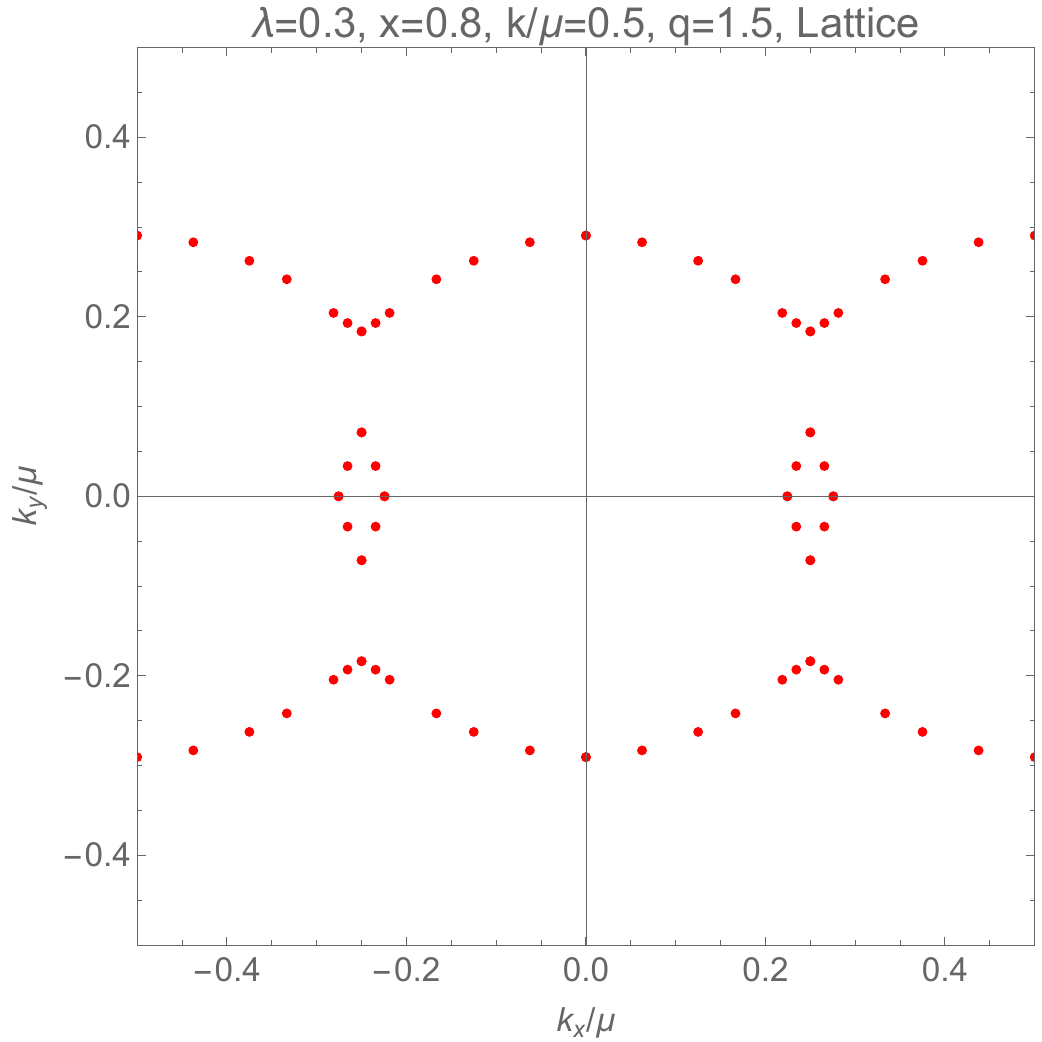}\ \hspace{0.8cm}
    \includegraphics[width=0.45\textwidth]{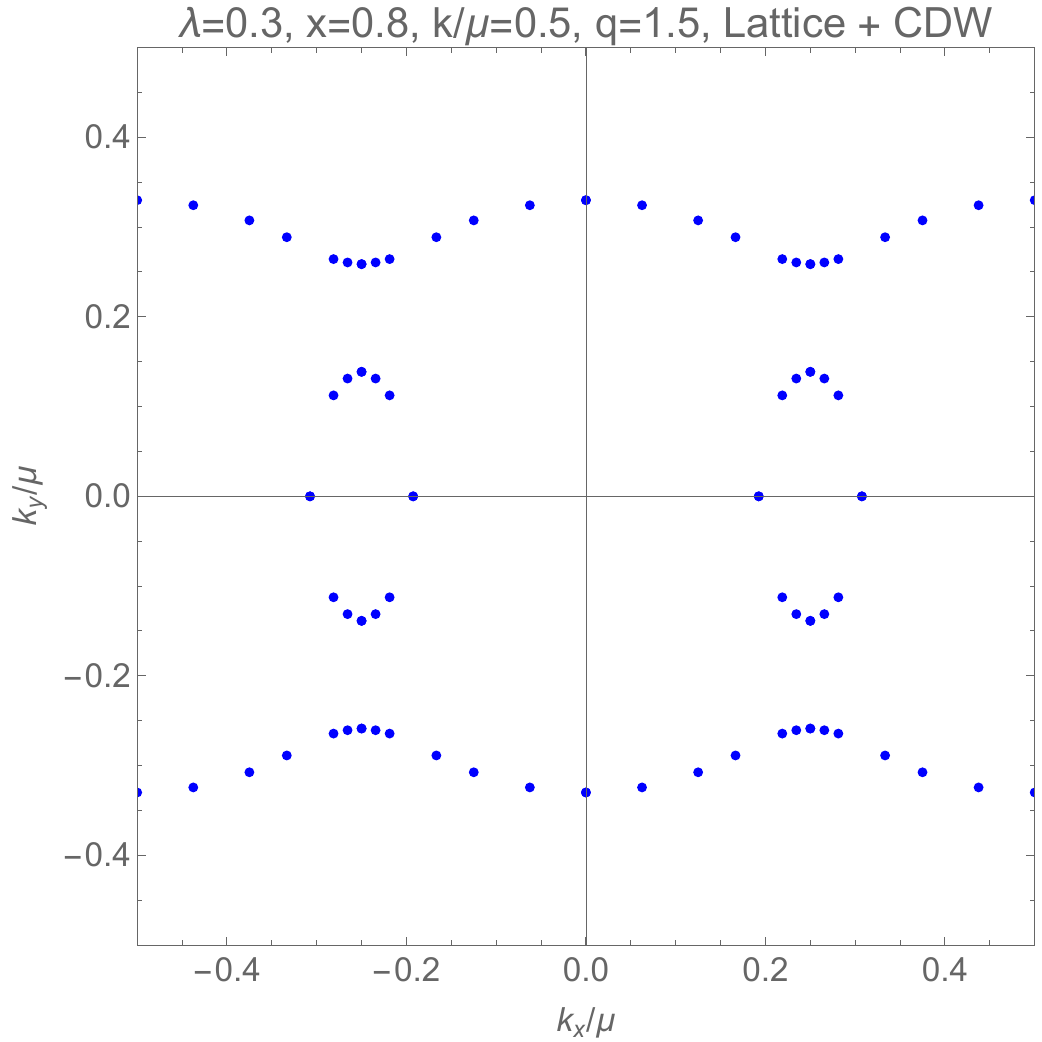}\ \hspace{0.05cm}
    \caption{\label{fig-4} \textbf{Top:} Fermi surface at $X=0.5$ (left), and $0.7$ (right), with fixed $\lambda=0.3$.
    \textbf{Bottom:} $X=0.8$ at the same $\lambda$, showing ionic lattice (left) versus lattice$+$CDW (right).}}
\end{figure}

\subsubsection{Formation of Band Gaps at Brillouin Zone Boundaries}

\begin{figure} [h]
  \center{
    \includegraphics[width=0.45\textwidth]{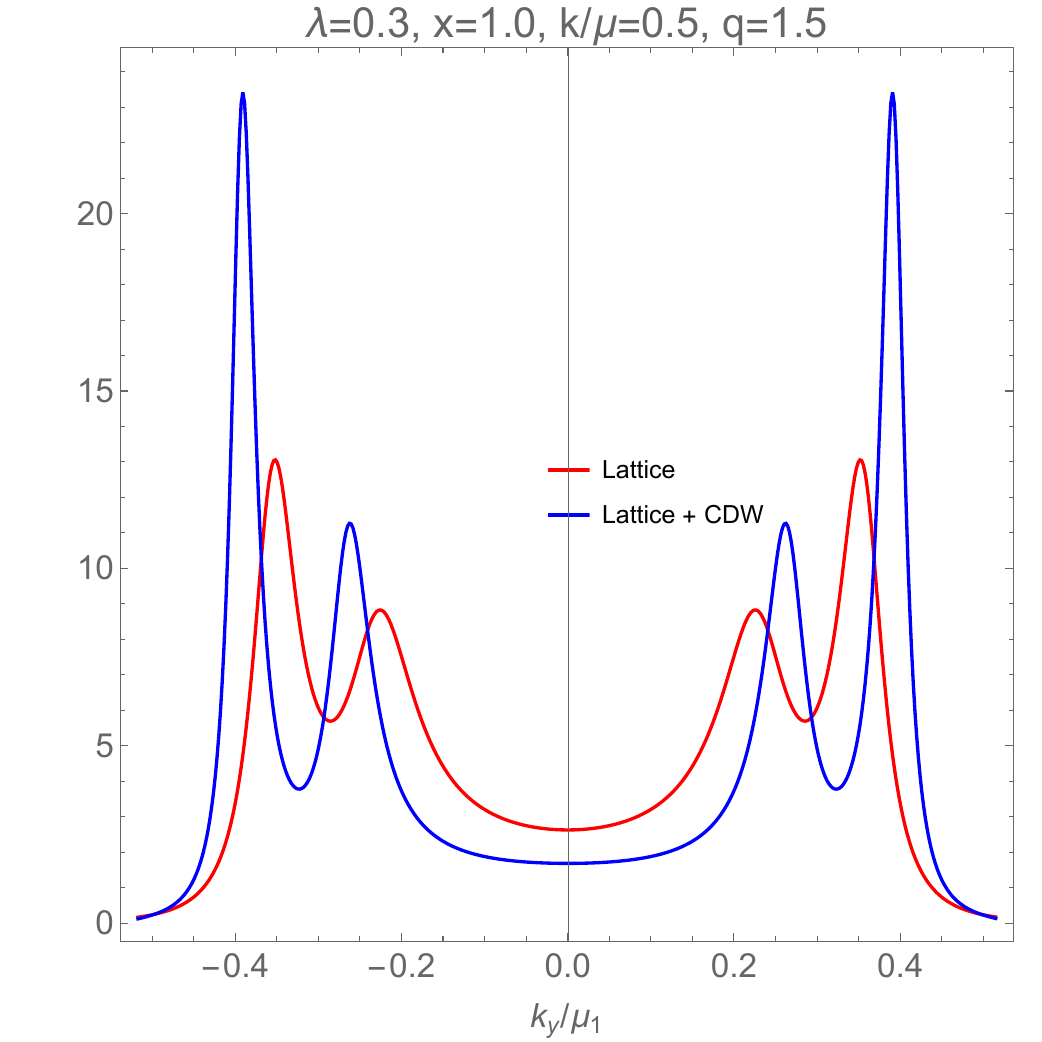}\ \hspace{0.8cm}
     \includegraphics[width=0.45\textwidth]{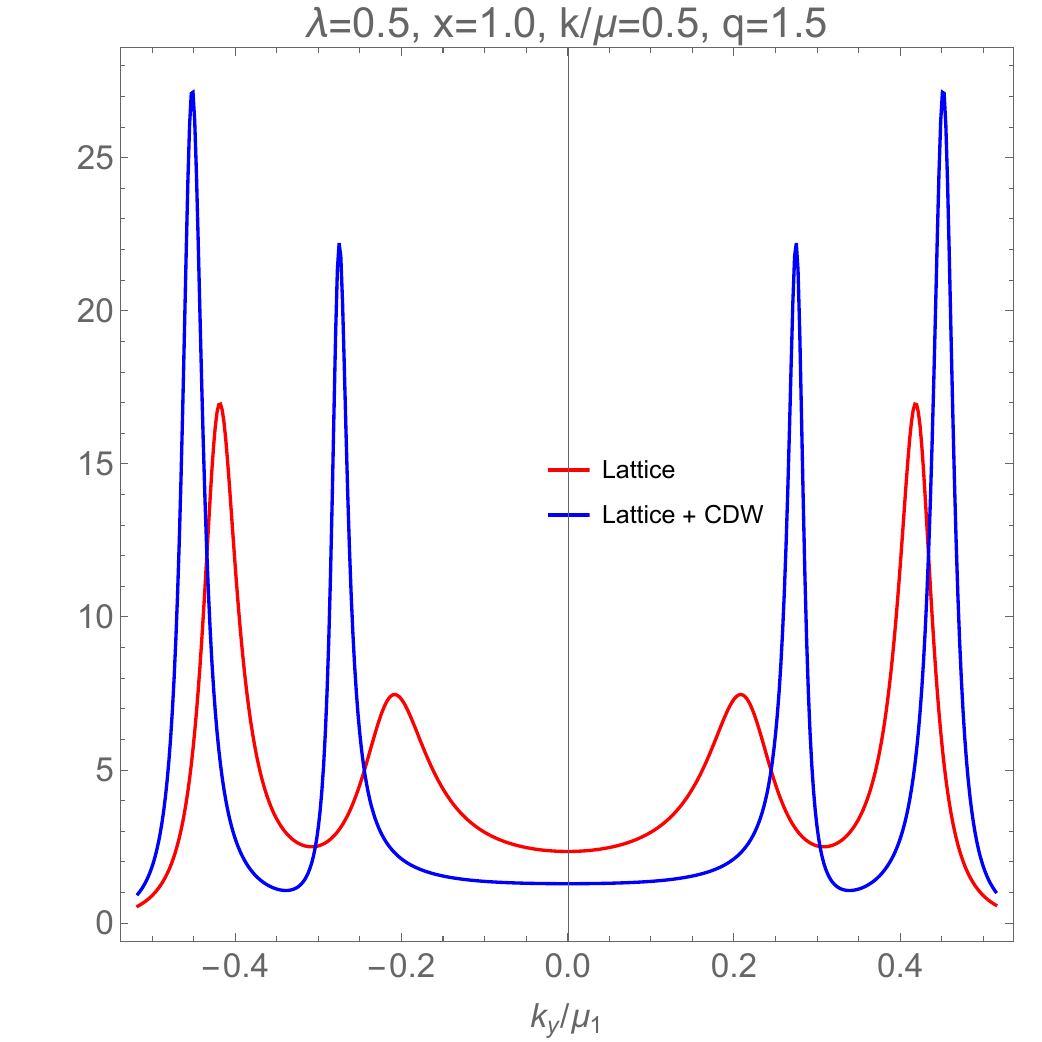}\ \hspace{0.8cm}
    \caption{\label{fig-5-1} The change of spectral function $A(\omega\!\to\!0,\mathbf{k})$ with the lattice amplitude, where we take $k_x/\mu_1=0.25$.}}
\end{figure}

\begin{figure} [h]
  \center{
     \includegraphics[width=0.45\textwidth]{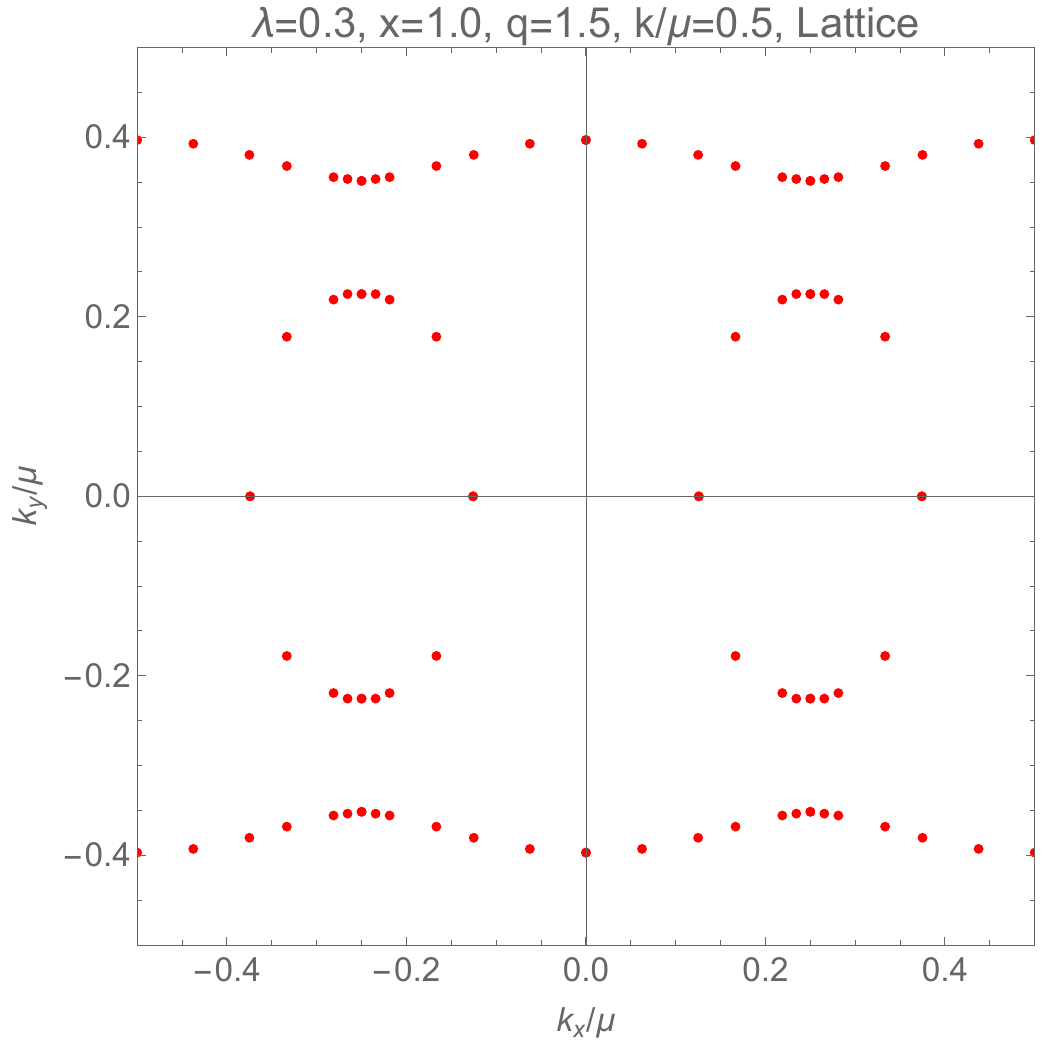}\ \hspace{0.8cm}
    \includegraphics[width=0.45\textwidth]{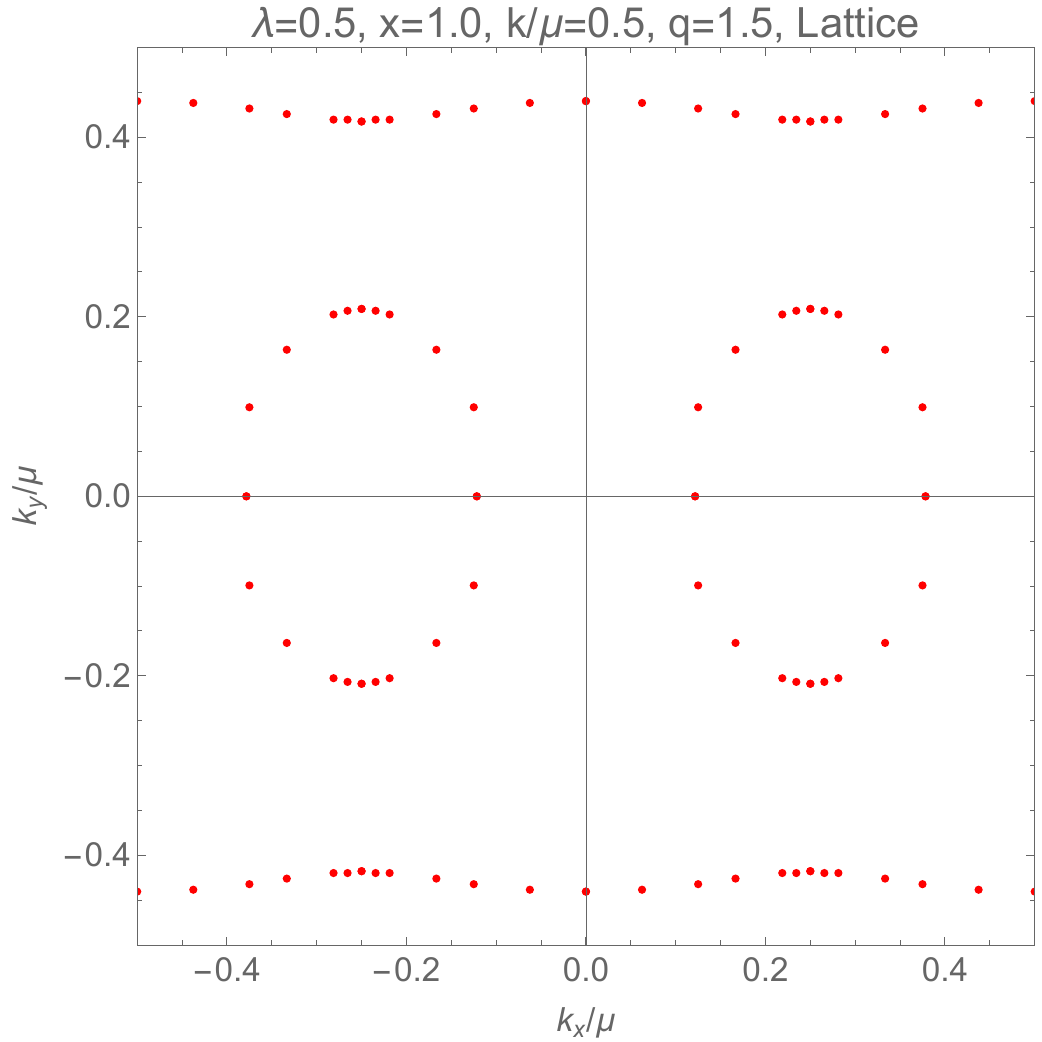}\ \hspace{0.0cm}
    \includegraphics[width=0.45\textwidth]{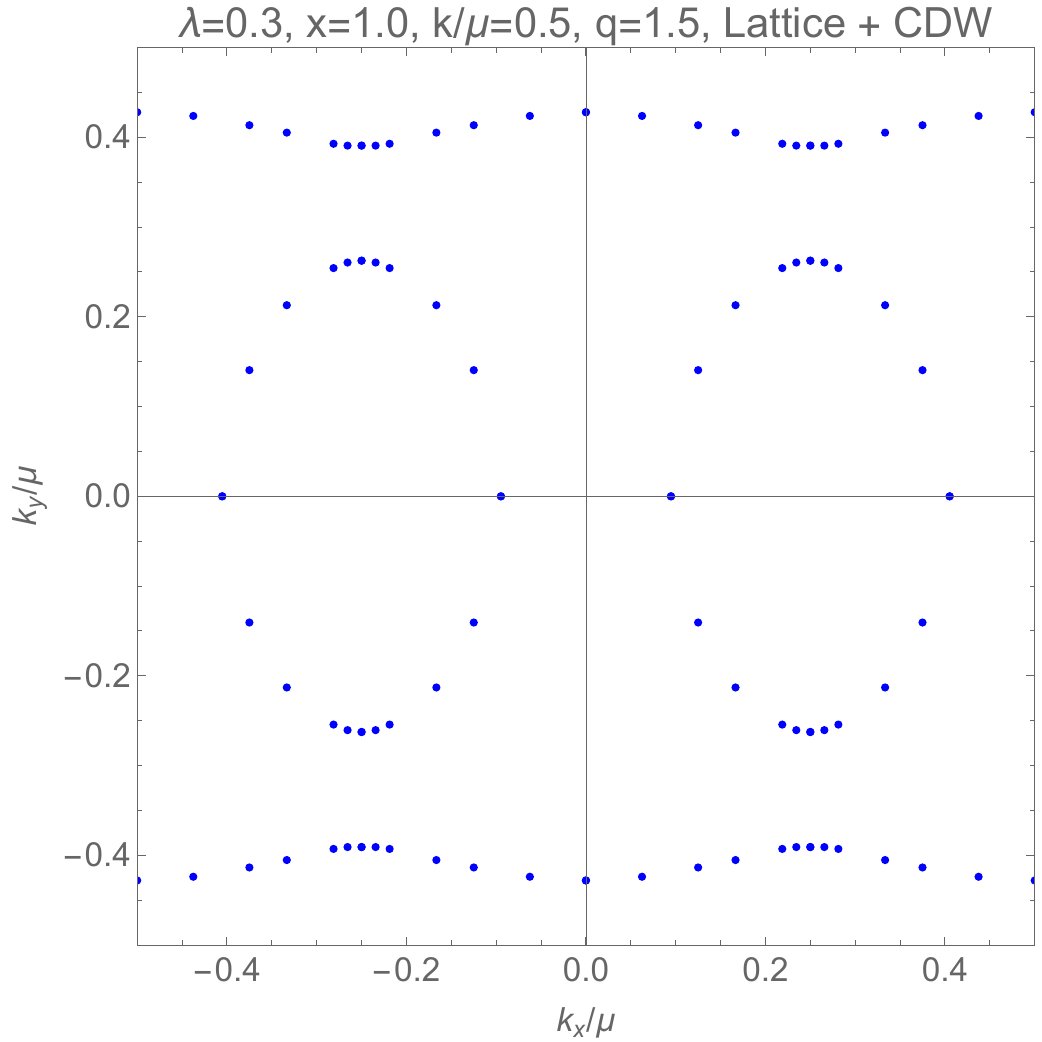}\ \hspace{0.8cm}
    \includegraphics[width=0.45\textwidth]{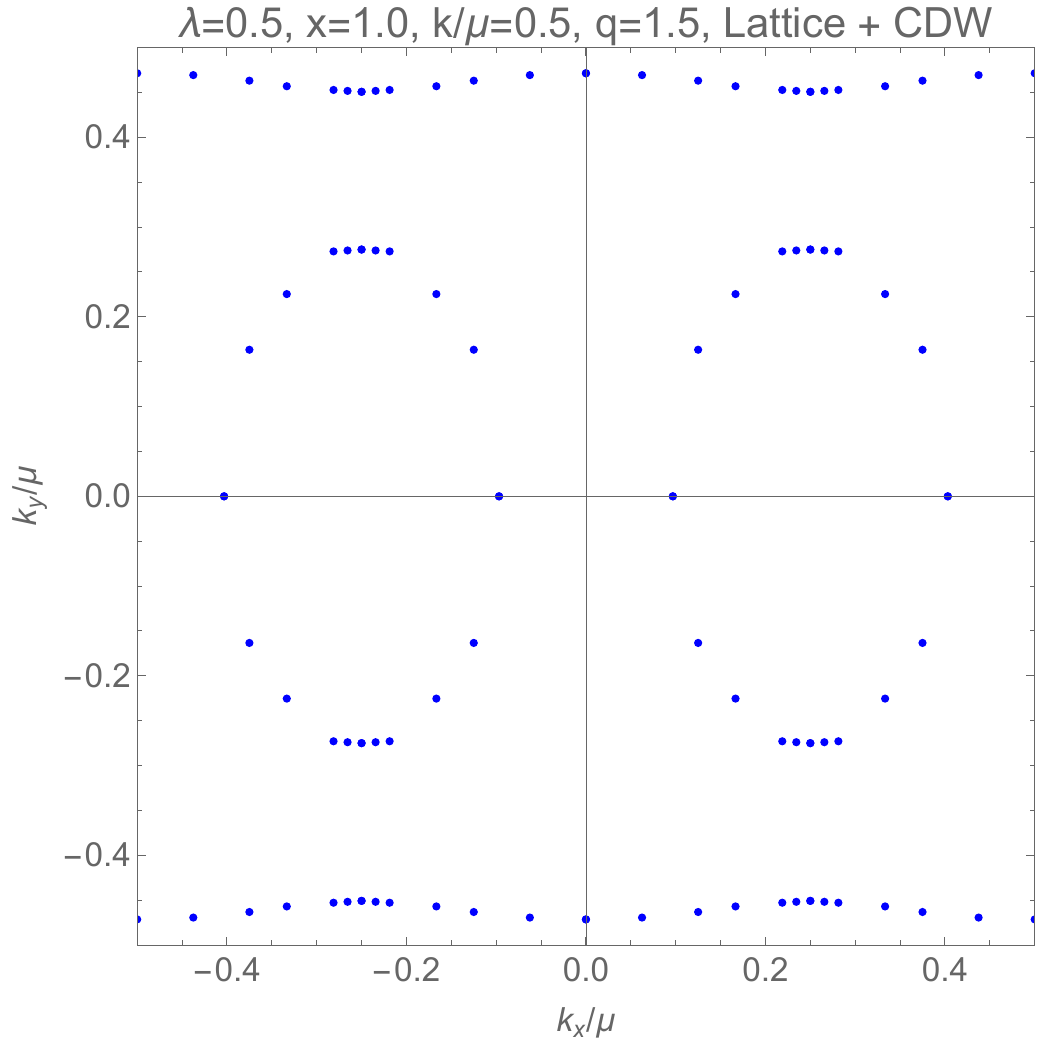}\ \hspace{0.0cm}
    \caption{\label{fig-5-2} The change of Fermi surface with the lattice amplitude.}
    }
\end{figure}

In holographic models, conventionally one could adjust the wave vector $k/\mu_1$ of the lattice to change the domain of the Brillouin zones such that the Fermi surface may cross the Brillouin zones. Alternatively, in our present model, we notice that the location of Fermi surface is sensitive to the doping parameter $X$ as well. Thus, we may observe the Fermi surface crossing the first Brillouin zone by increasing the doping parameter $X$ but fixing the wave vector of the lattice. An example is illustrated in Fig. \ref{fig-4}, where the lattice wave vector is fixed at $k/\mu_1=0.5$ such that the boundary of the first Brillouin zone lies at $k_x/\mu = \pm 0.25$. Therefore, when the Fermi momentum is larger than $k_x/\mu = 0.25$, then it is expected that the Fermi surface will cross the first Brillouin zone and a band gap at the boundary should appear due to Umklapp scattering. Mathematically, this corresponds to scattering between different Brillouin zones with a momentum transfer $\mathbf{K} = (K,0)$ where $K = 2\pi/L$ is the reciprocal lattice vector. Specifically, when the Fermi surface extends beyond the first Brillouin zone (i.e., when $k_F > K/2$), this Umklapp scattering becomes the dominant mechanism. It induces eigenvalue repulsion at degenerate energy levels, causing the spectral function to split into two peaks separated by a band gap, as illustrated in Fig. (\ref{fig-5-1}). It is worth noting that the strength of this gap opening is sensitive to the doping level. As discussed previously, the suppression of CDW order at higher doping levels weakens the Umklapp scattering channel across all momentum scales. This effect is most pronounced near the Brillouin zone boundaries, where it directly influences the magnitude of the opened gap.

\begin{table}[h]
  \centering
  \caption{The width of Band-gap at the Brillouin-zone boundary with $k/\mu_1=0.5$.\label{table-1}}
  \begin{tabular}{c@{\hspace{1.0cm}}c@{\hspace{1.0cm}}c@{\hspace{1.0cm}}c@{\hspace{1.0cm}}c@{\hspace{1.0cm}}c@{\hspace{0.5cm}}}
  \hline \hline & & $\text{X}=0.7$ & $\text{X}=0.8$ & $\text{X}=0.9$& $\text{X}=1.0$  \\
  \hline
 \multirow{2}{*}{$\lambda=$ 0.3} 
 & Lattice & $--$   & $0.120$   & $0.124$  & $0.126$ \\
 &  Lattice+CDW  & $0.116$   & $0.119$   & $0.122$  & $0.124$ \\
  \hline 
 \multirow{2}{*}{$\lambda=$ 0.5} 
 & Lattice & $--$   & $0.198$   & $0.202$  & $0.207$ \\
 &  Lattice+CDW  & $0.189$   & $0.195$   & $0.200$  & $0.204$ \\
   \hline 
 \multirow{2}{*}{$\lambda=$ 0.7} 
 & Lattice & $0.268$   & $0.271$   & $0.278$  & $0.283$ \\
 &  Lattice+CDW  & $0.259$   & $0.268$   & $0.275$  & $0.280$ \\
    \hline 
 \multirow{2}{*}{$\lambda=$ 1.0} 
 & Lattice & $0.369$   & $0.376$   & $0.383$  & $0.387$ \\
 &  Lattice+CDW  & $0.364$   & $0.374$   & $0.380$  & $0.382$ \\
     \hline 
 \multirow{2}{*}{$\lambda=$ 1.2} 
 & Lattice & $0.436$   & $0.443$   & $0.448$  & $0.447$ \\
 &  Lattice+CDW  & $0.433$   & $0.440$   & $0.443$  & $0.441$ \\
    \hline 
 \multirow{2}{*}{$\lambda=$ 1.5} 
 & Lattice & $0.534$   & $0.535$   & $0.530$  & $0.518$ \\
 &  Lattice+CDW  & $0.529$   & $0.529$   & $0.523$  & $0.510$ \\
  \hline \hline
  \end{tabular}
\end{table}

\begin{figure} [h]
  \center{
    \includegraphics[width=0.7\textwidth]{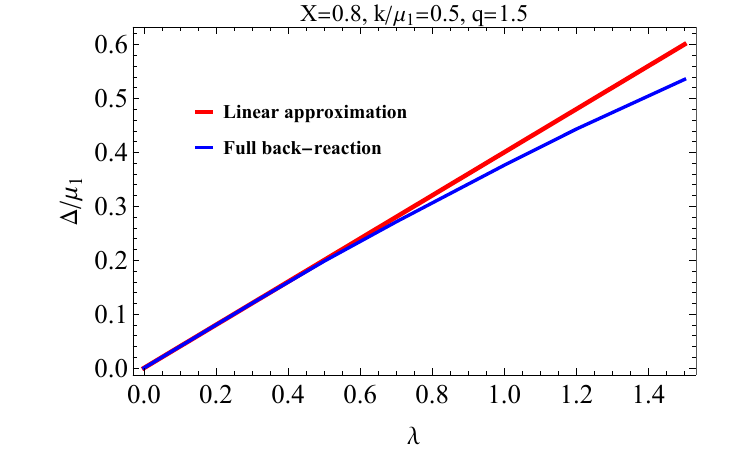}
    \caption{\label{fig-gap-lambda}
    The width of band-gap $\Delta/\mu_1$ versus the lattice amplitude $\lambda$ when other parameters are fixed.}}
\end{figure}

Interesting enough, Fig. \ref{fig-4} manifestly demonstrates the process of the Fermi surface crossing the first Brillouin zone with the increase of the doping parameter $X$. In particular, after crossing the Brillouin zone, the band gap is obviously observed at the boundary as the separation of two peaks with the same $k_x/\mu_1=0.25$.

The correspondence between the change of spectral function $A(\omega\!\to\!0,\mathbf{k})$ and Fermi surface with the change of the lattice amplitude $\lambda$ is illustrated in Fig. \ref{fig-5-1} and Fig. \ref{fig-5-2}, with a fixed doping parameter $X=1.0$. These figures systematically demonstrate how the explicit symmetry breaking strength modulates the electronic structure. In Fig. \ref{fig-5-1}, we plot the spectral function as a function of $k_y$  at the first Brillouin zone boundary $k_x/\mu_1 = 0.25$ for different lattice amplitudes. As $\lambda$ increases, two peaks in the spectral function become increasingly separated, with the growing separation distance quantifying the widening band gap $\Delta$. This progressive gap enhancement directly reflects the strengthening Umklapp scattering at zone boundaries. Meanwhile, Fig. \ref{fig-5-2} demonstrates the variation of the Fermi surface with increasing lattice amplitude. Notably, comparing the lattice-only and lattice+CDW cases, we observe that CDW-induced charge screening partially compensates the ionic lattice potential, resulting in systematically smaller band gaps across all parameter regimes, as quantitatively shown in Fig. \ref{fig-5-1} and Table \ref{table-1}. Evidently, the band gap increases with the increase of the lattice amplitude in both lattice case and lattice+CDW case, demonstrating the non-additive nature of combined symmetry breaking mechanisms.

\subsubsection{Interplay Between Lattice and CDW Effects on Band Gap Structure}
One of the most intriguing findings in our study concerns the interplay between the ionic lattice and CDW in determining the band gap structure. Our numerical results are presented in Table \ref{table-1} and we summarize the rules observed in this table as the following list.
\begin{enumerate}
    \item The width of band gap versus the doping parameter $X$. For moderate lattice amplitudes, the band gap increases with $X$ in both lattice and lattice+CDW cases, reflecting enhanced Fermi surface warping near the Brillouin zone boundaries where Umklapp scattering is most effective. However, at larger lattice amplitudes ($\lambda \gtrsim 1.5$), competing effects lead to non-monotonic behavior, with the band gap decreasing at high doping levels due to enhanced screening and topological modifications of the electronic structure.
    \item  The width of band gap versus the lattice amplitude  $\lambda$. Obviously, the width of the band gap becomes larger with the increase of the lattice amplitude in both lattice and lattice+CDW cases. It is also quite understandable since the lattice effect causes the formation of the band gap and it becomes stronger with the larger amplitude. To quantitatively describe this relation we plot Figure. \ref{fig-gap-lambda} for $\Delta/\mu_1$ versus $\lambda$ with other parameter fixed. We notice that with small lattice amplitude, the width of the band gap increases with the amplitude almost linearly, while for larger $\lambda$ the deviation from linearity  becomes evident.  
    \item The effects of CDW on the width of band gap. As shown in Table \ref{table-1}, the band gap in the lattice+CDW system is systematically smaller than in the pure lattice case. This narrowing arises from CDW-induced charge screening: when CDW forms spontaneously, the charge redistribution 
\begin{equation}
\Delta\rho(x) = \rho_{\text{lattice+CDW}}(x) - \rho_{\text{lattice-only}}(x)
\end{equation}
tends to accumulate where the lattice potential is most negative, thereby partially compensating its spatial variation. This screening reduces Umklapp scattering at Brillouin zone boundaries, resulting in smaller gaps. 
\end{enumerate}
 
Our findings reveal important differences from previous holographic studies. Ref. \cite{Mukhopadhyay:2020tky} found that introducing ionic lattices widens the gap at Brillouin zone boundaries in D2-D8 CDW model. Their study demonstrated gap enhancement when adding lattice to an existing CDW background. In contrast, our system—featuring both ionic lattices and CDW from the outset—exhibits gap reduction when comparing lattice+CDW with lattice-only configurations at identical parameters. This non-additive interaction suggests that the sequence of introducing symmetry-breaking mechanisms matters: adding lattice to existing CDW (their approach) versus having both coexist initially (our setup) leads to qualitatively different outcomes. This distinction provides new insights into how the relative timing and strength of multiple ordering tendencies influence electronic band structures in strongly correlated materials.

\subsection{The Commensurability Effects on the Fermi Surface \label{subsectC}}

\begin{figure} [h]
  \center{
     \includegraphics[width=0.6\textwidth]{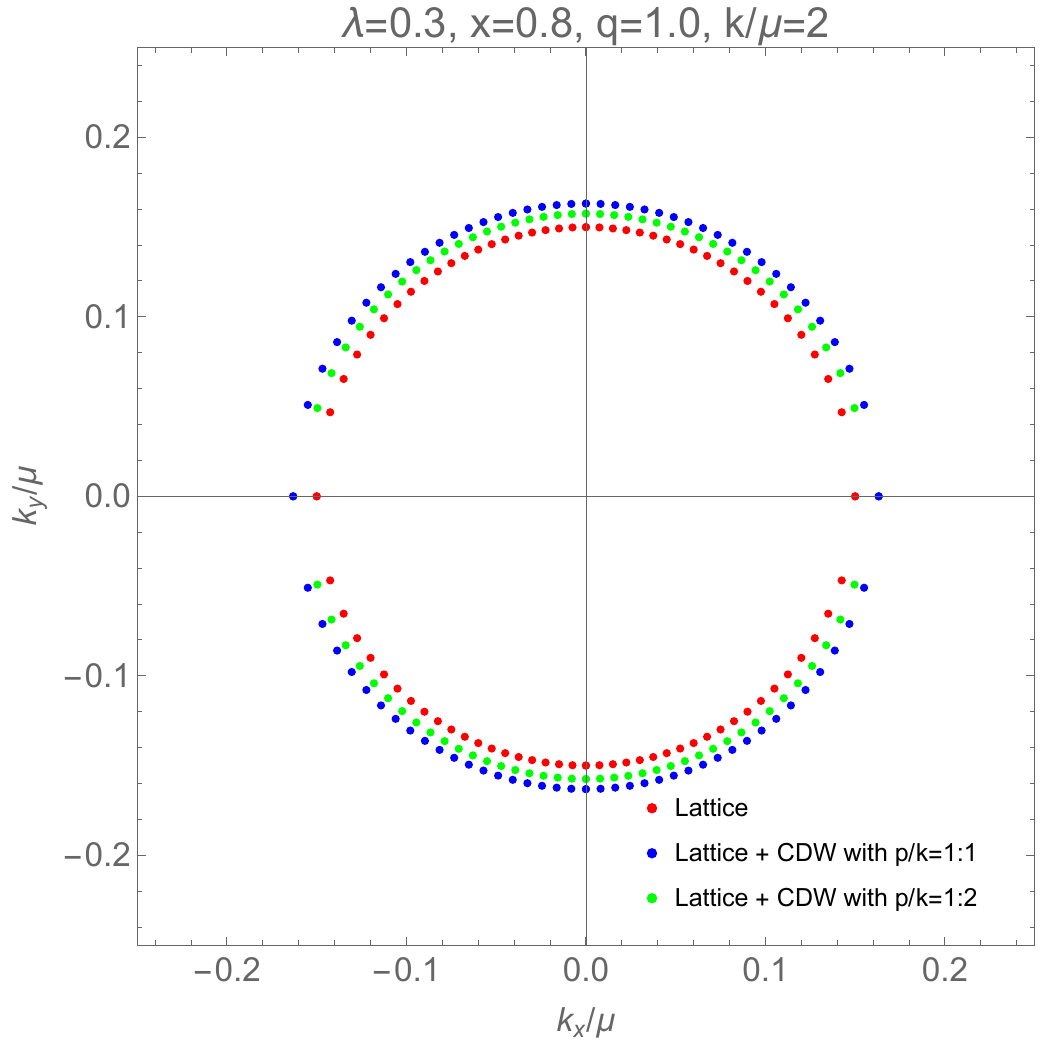}\ \hspace{0.05cm}
    \caption{\label{fig-8}  The Fermi surface
    for the lattice only (red) and lattice$+$CDW at different commensurate ratios: $p/k = 1:1$ (blue) and $p/k = 1:2$ (green).}}
\end{figure}

The relationship between the commensurability of the CDW wave vector and ionic lattice periodicity as well as the resulting fermionic properties in our holographic system exhibits interesting behavior that warrants investigation. 

\subsubsection{Commensurate ratio dependence}
The commensurate ratio between the CDW wave vector and the ionic lattice periodicity imposes strict geometric constraints on the fermionic spectrum. Fig. \ref{fig-8} contrasts the Fermi surface reconstruction under $p/k = 1:1$ and $p/k = 1:2$ configurations against the lattice-only background. The $p/k = 1:2$ configuration creates a superlattice potential that induces a significantly smaller Fermi surface volume compared to the $1:1$ case.

This suppression originates from the distinct scattering channels allowed by the commensurate geometry. The $1:2$ ratio introduces an additional periodicity twice that of the underlying lattice, effectively halving the size of the Brillouin zone and enabling new Umklapp scattering processes at $G = \pm k/2$. These additional scattering channels enhance the quasiparticle backscattering, leading to a stronger confinement of the Fermi surface. Furthermore, as indicated by the thermodynamic analysis in Ref.~\cite{Li:2024ybq}, the Type~II CDW naturally locks in at $p/k = 1:2$ with a lower free energy state. This thermodynamic stability implies that the $1:2$ configuration provides a robust potential landscape, thereby maximizing the spectral weight redistribution and facilitating the observed Fermi surface reduction.

\begin{figure} [h]
\center{
  \includegraphics[width=0.49\textwidth]{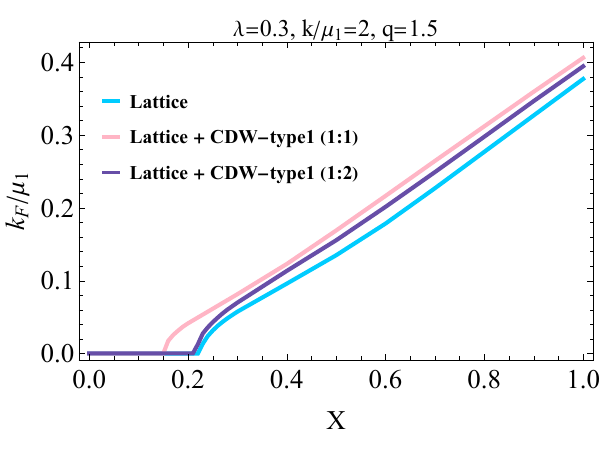}\ \hspace{0.05cm}
  \includegraphics[width=0.49\textwidth]{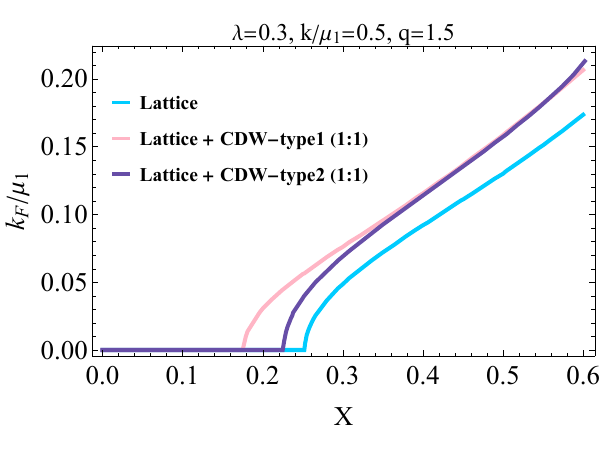}
\caption{\label{fig-6} Fermi momentum $k_F/\mu_1$ versus $X$ for different $k/\mu_1$ and $q$.}}
\end{figure}

\subsubsection{CDW type dependence at fixed commensurate ratio}
When the commensurate ratio is fixed at $p/k = 1:1$, the fermionic response becomes sensitive to the specific internal structure of the CDW solution. Fig. \ref{fig-6} illustrates the systematic variation of the Fermi momentum $k_F$ with doping $X$, revealing a quantitative discrepancy between Type~I and Type~II solutions (right panel).

This difference is fundamentally driven by the distinct charge density profiles of the two types. The Type~I solution is characterized by a non-negligible DC component ($\Delta \rho^{(0)}$) in its charge modulation, which is absent or negligible in Type~II. According to the Luttinger theorem, this DC component effectively renormalizes the background carrier density, acting as a ``local doping" shift that rigidly expands or contracts the Fermi surface volume. Consequently, Type~I exhibits a marked deviation from the lattice-only baseline, particularly at low doping ($k/\mu_1 = 0.5$, left panel) where the CDW order is strongest. As doping $X$ increases ($k/\mu_1 = 2$, right panel), the enhanced screening from mobile carriers suppresses the CDW amplitude. This screening effect diminishes the magnitude of the DC component, causing the Fermi momenta of Type~I and Type~II to converge, thereby confirming that the distinct fermionic signatures are dominated by the interplay between the CDW-induced background renormalization and the metallic screening.

\subsubsection{Unified physical picture and experimental connections}
The phenomenological richness of the system arises from the tripartite interplay governing charge redistribution: geometric matching (commensurate ratio), background charge renormalization (CDW type and DC component), and metallic screening (doping level). 

A clear stability hierarchy emerges: the $p/k = 1:2$ configuration, which is thermodynamically favored for Type~II solutions~\cite{Li:2024ybq}, exhibits a stronger lock-in effect than the $1:1$ case, creating a robust superlattice potential that significantly reconstructs the Fermi surface. The doping evolution—transitioning from a CDW-dominated regime with strong spectral weight suppression at low $X$ to a coherent itinerant regime at high $X$—provides a holographic analog of the crossover from the pseudogap phase to the Fermi liquid in cuprates \cite{boebinger1996insulator,Vojta01112009}. Our analysis predicts that commensurate lock-in effects should manifest experimentally as distinct spectral weight suppression or gap opening at specific momenta in ARPES measurements. Crucially, these signatures are predicted to be most distinguishable at intermediate doping, where the competition between the geometric constraints of the ionic lattice and the screening by mobile carriers is most intense.

\section{Discussion}\label{sec:dis}
In this work, we have investigated the holographic fermions over a background that contains an ionic lattice due to the explicit breaking of translational symmetry and charge density waves due to the spontaneous breaking of translational symmetry. Our results reveal significant modifications to the structure of the Fermi surface and the formation of the band gap compared to the homogeneous cases. The interplay between these two mechanisms of symmetry breaking leads to distinctive features in the fermionic spectral function, particularly at the Brillouin zone boundaries where we observe clear band gap formation. These results align with previous studies, but extend the analysis to incorporate the crucial element of ionic lattices that more accurately reflect real condensed matter systems.

The formation and variation of the Fermi surface in our model demonstrate several key physical insights. First, the presence of the ionic lattice as well as the CDW order significantly affects the shape and the coherence of the Fermi surface. It is found that the formation of
the CDW enhances the amplitude of spectral function as well as the momentum of the Fermi surface. Secondly, we are concerned with the change of the band gap with the doping parameter as well as the lattice amplitude. When the Fermi surface intersects with the Brillouin zone boundary, we observe the development of band gaps whose magnitude depends on both the strength of the ionic lattice and the amplitude of the CDW. Interestingly, we find that the radius of the Fermi
surface expands with the increase of the doping parameter and finally may cross the first Brillouin
zone. Notably, our results reveal that the presence of CDW leads to systematically smaller band gaps compared to the pure lattice case, indicating a screening mechanism where CDW-induced charge redistribution partially compensates the ionic lattice potential. This non-additive interaction between explicit and spontaneous symmetry breaking provides a holographic realization of the complex band structure modifications that occur in strongly correlated electron systems exhibiting density wave phases. Furthermore, our findings demonstrate that the sequence of introducing different symmetry-breaking mechanisms matters: systems where lattice and CDW coexist from the outset exhibit qualitatively different behavior compared to scenarios where one mechanism is added to an existing background. These insights offer potential connections to materials such as cuprate superconductors where both lattice effects and spontaneous ordering are present.

Our observation of band gap reduction in the lattice+CDW system, rather than enhancement, highlights the importance of carefully considering the interplay mechanisms in systems with multiple competing orders. The screening effect we identify—where spontaneous CDW formation redistributes charge to partially compensate the explicit lattice potential—represents a physically distinct mechanism from simple additive effects. This finding suggests that in real materials, the coexistence of multiple ordering tendencies may lead to unexpected modifications of electronic properties that cannot be predicted by considering each mechanism in isolation.

Looking forward, several directions for future research emerge from our work. For simplicity, the possible couplings between the fermions and other fields in the background are not taken into account at current stage, such as the coupling between the scalar field and the fermions and the dipole coupling between the gauge field and the fermions \cite{Faulkner:2009am,Edalati:2010ge,Edalati:2010ww,Chakrabarti:2021qie,Wahlang:2021oqk,Chakrabarti:2022lxl}. Such couplings would play a significant role in exploring the structure of the energy gap in the system. Another natural extension would be investigating the temperature dependence of the spectral function, particularly near the critical temperature where the CDW order begins to form. Additionally, incorporating superconducting order into our model would allow for a more comprehensive study of the intertwined orders observed in high-temperature superconductors, where PDW (pair density wave) order coexists with CDW. As suggested by the recent work \cite{Li:2024ybq} on holographic striped superconductors with ionic lattices, exploring the commensurate lock-in effect in systems with multiple competing orders could provide valuable insights into the phase diagram of high-temperature superconductivity. Finally, a detailed analysis of transport properties in this background, particularly the optical conductivity, would establish connections between our theoretical predictions and experimentally observable quantities, potentially offering new perspectives on non-Fermi liquid behavior in strongly correlated systems.

\section*{Acknowledgments}
We are very grateful to Fang-jing Cheng, Bin-ye Dong, Zhangping Yu for helpful discussions. This work is supported in part by the Natural Science Foundation of China (Grant Nos.~12275275,~12475054,~12405067), the Guangdong Basic and Applied Basic Research Foundation No. 2025A1515012063 and the Natural Science Foundation of Sichuan (No. 2025ZNSFSC0876).

\bibliographystyle{style1}% common bib file
\bibliography{scalar}

\providecommand{\href}[2]{#2}\begingroup\raggedright\begin{thebibliography}{10}

\bibitem{Maldacena:1997re}
J.~M. Maldacena, {\it {The Large $N$ limit of superconformal field theories and supergravity}},  {\em Adv. Theor. Math. Phys.} {\bf 2} (1998) 231--252, [\href{http://arxiv.org/abs/hep-th/9711200}{{\tt hep-th/9711200}}].

\bibitem{Gubser:1998bc}
S.~S. Gubser, I.~R. Klebanov, and A.~M. Polyakov, {\it {Gauge theory correlators from noncritical string theory}},  {\em Phys. Lett. B} {\bf 428} (1998) 105--114, [\href{http://arxiv.org/abs/hep-th/9802109}{{\tt hep-th/9802109}}].

\bibitem{Witten:1998qj}
E.~Witten, {\it {Anti de Sitter space and holography}},  {\em Adv. Theor. Math. Phys.} {\bf 2} (1998) 253--291, [\href{http://arxiv.org/abs/hep-th/9802150}{{\tt hep-th/9802150}}].

\bibitem{Hartnoll:2009sz}
S.~A. Hartnoll, {\it {Lectures on holographic methods for condensed matter physics}},  {\em Class. Quant. Grav.} {\bf 26} (2009) 224002, [\href{http://arxiv.org/abs/0903.3246}{{\tt arXiv:0903.3246}}].

\bibitem{Iqbal:2011ae}
N.~Iqbal, H.~Liu, and M.~Mezei, {\it {Lectures on holographic non-Fermi liquids and quantum phase transitions}},  in {\em {Theoretical Advanced Study Institute in Elementary Particle Physics}: {String theory and its Applications: From meV to the Planck Scale}}, pp.~707--816, 10, 2011.
\newblock \href{http://arxiv.org/abs/1110.3814}{{\tt arXiv:1110.3814}}.

\bibitem{Hartnoll:2016apf}
S.~A. Hartnoll, A.~Lucas, and S.~Sachdev, {\it {Holographic quantum matter}},  \href{http://arxiv.org/abs/1612.07324}{{\tt arXiv:1612.07324}}.

\bibitem{Zaanen:2015oix}
J.~Zaanen, Y.-W. Sun, Y.~Liu, and K.~Schalm, {\em {Holographic Duality in Condensed Matter Physics}}.
\newblock Cambridge Univ. Press, 2015.

\bibitem{Ammon:2015wua}
M.~Ammon and J.~Erdmenger, {\em {Gauge/gravity duality}: {Foundations and applications}}.
\newblock Cambridge University Press, Cambridge, 4, 2015.

\bibitem{Gubser:2008px}
S.~S. Gubser, {\it {Breaking an Abelian gauge symmetry near a black hole horizon}},  {\em Phys. Rev. D} {\bf 78} (2008) 065034, [\href{http://arxiv.org/abs/0801.2977}{{\tt arXiv:0801.2977}}].

\bibitem{Hartnoll:2008vx}
S.~A. Hartnoll, C.~P. Herzog, and G.~T. Horowitz, {\it {Building a Holographic Superconductor}},  {\em Phys. Rev. Lett.} {\bf 101} (2008) 031601, [\href{http://arxiv.org/abs/0803.3295}{{\tt arXiv:0803.3295}}].

\bibitem{Hartnoll:2008kx}
S.~A. Hartnoll, C.~P. Herzog, and G.~T. Horowitz, {\it {Holographic Superconductors}},  {\em JHEP} {\bf 12} (2008) 015, [\href{http://arxiv.org/abs/0810.1563}{{\tt arXiv:0810.1563}}].

\bibitem{Hartnoll:2009ns}
S.~A. Hartnoll, J.~Polchinski, E.~Silverstein, and D.~Tong, {\it {Towards strange metallic holography}},  {\em JHEP} {\bf 04} (2010) 120, [\href{http://arxiv.org/abs/0912.1061}{{\tt arXiv:0912.1061}}].

\bibitem{Faulkner:2010zz}
T.~Faulkner, N.~Iqbal, H.~Liu, J.~McGreevy, and D.~Vegh, {\it {Strange metal transport realized by gauge/gravity duality}},  {\em Science} {\bf 329} (2010) 1043--1047.

\bibitem{Cubrovic:2009ye}
M.~Cubrovic, J.~Zaanen, and K.~Schalm, {\it {String Theory, Quantum Phase Transitions and the Emergent Fermi-Liquid}},  {\em Science} {\bf 325} (2009) 439--444, [\href{http://arxiv.org/abs/0904.1993}{{\tt arXiv:0904.1993}}].

\bibitem{Hartnoll:2012rj}
S.~A. Hartnoll and D.~M. Hofman, {\it {Locally Critical Resistivities from Umklapp Scattering}},  {\em Phys. Rev. Lett.} {\bf 108} (2012) 241601, [\href{http://arxiv.org/abs/1201.3917}{{\tt arXiv:1201.3917}}].

\bibitem{Horowitz:2012ky}
G.~T. Horowitz, J.~E. Santos, and D.~Tong, {\it {Optical Conductivity with Holographic Lattices}},  {\em JHEP} {\bf 07} (2012) 168, [\href{http://arxiv.org/abs/1204.0519}{{\tt arXiv:1204.0519}}].

\bibitem{Horowitz:2012gs}
G.~T. Horowitz, J.~E. Santos, and D.~Tong, {\it {Further Evidence for Lattice-Induced Scaling}},  {\em JHEP} {\bf 11} (2012) 102, [\href{http://arxiv.org/abs/1209.1098}{{\tt arXiv:1209.1098}}].

\bibitem{Horowitz:2013jaa}
G.~T. Horowitz and J.~E. Santos, {\it {General Relativity and the Cuprates}},  {\em JHEP} {\bf 06} (2013) 087, [\href{http://arxiv.org/abs/1302.6586}{{\tt arXiv:1302.6586}}].

\bibitem{Ling:2013nxa}
Y.~Ling, C.~Niu, J.-P. Wu, and Z.-Y. Xian, {\it {Holographic Lattice in Einstein-Maxwell-Dilaton Gravity}},  {\em JHEP} {\bf 11} (2013) 006, [\href{http://arxiv.org/abs/1309.4580}{{\tt arXiv:1309.4580}}].

\bibitem{Andrade:2013gsa}
T.~Andrade and B.~Withers, {\it {A simple holographic model of momentum relaxation}},  {\em JHEP} {\bf 05} (2014) 101, [\href{http://arxiv.org/abs/1311.5157}{{\tt arXiv:1311.5157}}].

\bibitem{Donos:2011bh}
A.~Donos and J.~P. Gauntlett, {\it {Holographic striped phases}},  {\em JHEP} {\bf 08} (2011) 140, [\href{http://arxiv.org/abs/1106.2004}{{\tt arXiv:1106.2004}}].

\bibitem{Withers:2013loa}
B.~Withers, {\it {Black branes dual to striped phases}},  {\em Class. Quant. Grav.} {\bf 30} (2013) 155025, [\href{http://arxiv.org/abs/1304.0129}{{\tt arXiv:1304.0129}}].

\bibitem{Donos:2013gda}
A.~Donos and J.~P. Gauntlett, {\it {Holographic charge density waves}},  {\em Phys. Rev. D} {\bf 87} (2013), no.~12 126008, [\href{http://arxiv.org/abs/1303.4398}{{\tt arXiv:1303.4398}}].

\bibitem{Donos:2013wia}
A.~Donos, {\it {Striped phases from holography}},  {\em JHEP} {\bf 05} (2013) 059, [\href{http://arxiv.org/abs/1303.7211}{{\tt arXiv:1303.7211}}].

\bibitem{Withers:2013kva}
B.~Withers, {\it {The moduli space of striped black branes}},  \href{http://arxiv.org/abs/1304.2011}{{\tt arXiv:1304.2011}}.

\bibitem{Ling:2014saa}
Y.~Ling, C.~Niu, J.~Wu, Z.~Xian, and H.-b. Zhang, {\it {Metal-insulator Transition by Holographic Charge Density Waves}},  {\em Phys. Rev. Lett.} {\bf 113} (2014) 091602, [\href{http://arxiv.org/abs/1404.0777}{{\tt arXiv:1404.0777}}].

\bibitem{Andrade:2015iyf}
T.~Andrade and A.~Krikun, {\it {Commensurability effects in holographic homogeneous lattices}},  {\em JHEP} {\bf 05} (2016) 039, [\href{http://arxiv.org/abs/1512.02465}{{\tt arXiv:1512.02465}}].

\bibitem{Baggioli:2016oju}
M.~Baggioli and O.~Pujolas, {\it {On Effective Holographic Mott Insulators}},  {\em JHEP} {\bf 12} (2016) 107, [\href{http://arxiv.org/abs/1604.08915}{{\tt arXiv:1604.08915}}].

\bibitem{Jokela:2017ltu}
N.~Jokela, M.~Jarvinen, and M.~Lippert, {\it {Pinning of holographic sliding stripes}},  {\em Phys. Rev. D} {\bf 96} (2017), no.~10 106017, [\href{http://arxiv.org/abs/1708.07837}{{\tt arXiv:1708.07837}}].

\bibitem{Krikun:2017cyw}
A.~Krikun, {\it {Holographic discommensurations}},  {\em JHEP} {\bf 12} (2018) 030, [\href{http://arxiv.org/abs/1710.05801}{{\tt arXiv:1710.05801}}].

\bibitem{Andrade:2017leb}
T.~Andrade and A.~Krikun, {\it {Commensurate lock-in in holographic non-homogeneous lattices}},  {\em JHEP} {\bf 03} (2017) 168, [\href{http://arxiv.org/abs/1701.04625}{{\tt arXiv:1701.04625}}].

\bibitem{Andrade:2017ghg}
T.~Andrade, A.~Krikun, K.~Schalm, and J.~Zaanen, {\it {Doping the holographic Mott insulator}},  {\em Nature Phys.} {\bf 14} (2018), no.~10 1049--1055, [\href{http://arxiv.org/abs/1710.05791}{{\tt arXiv:1710.05791}}].

\bibitem{Ammon:2019wci}
M.~Ammon, M.~Baggioli, and A.~Jim{\'e}nez-Alba, {\it {A Unified Description of Translational Symmetry Breaking in Holography}},  {\em JHEP} {\bf 09} (2019) 124, [\href{http://arxiv.org/abs/1904.05785}{{\tt arXiv:1904.05785}}].

\bibitem{Andrade:2020hpu}
T.~Andrade, M.~Baggioli, and A.~Krikun, {\it {Phase relaxation and pattern formation in holographic gapless charge density waves}},  {\em JHEP} {\bf 03} (2021) 292, [\href{http://arxiv.org/abs/2009.05551}{{\tt arXiv:2009.05551}}].

\bibitem{Baggioli:2021xuv}
M.~Baggioli, K.-Y. Kim, L.~Li, and W.-J. Li, {\it {Holographic Axion Model: a simple gravitational tool for quantum matter}},  {\em Sci. China Phys. Mech. Astron.} {\bf 64} (2021), no.~7 270001, [\href{http://arxiv.org/abs/2101.01892}{{\tt arXiv:2101.01892}}].

\bibitem{Baggioli:2022pyb}
M.~Baggioli and B.~Gout{\'e}raux, {\it {Colloquium: Hydrodynamics and holography of charge density wave phases}},  {\em Rev. Mod. Phys.} {\bf 95} (2023), no.~1 011001, [\href{http://arxiv.org/abs/2203.03298}{{\tt arXiv:2203.03298}}].

\bibitem{Ling:2020qdd}
Y.~Ling and M.-H. Wu, {\it {Holographic striped superconductor}},  {\em JHEP} {\bf 03} (2021) 260, [\href{http://arxiv.org/abs/2011.12150}{{\tt arXiv:2011.12150}}].

\bibitem{Li:2024ybq}
K.~Li, Y.~Ling, P.~Liu, and M.-H. Wu, {\it {Holographic striped superconductor with ionic lattice}},  {\em JHEP} {\bf 02} (2025) 028, [\href{http://arxiv.org/abs/2411.10181}{{\tt arXiv:2411.10181}}].

\bibitem{Faulkner:2009wj}
T.~Faulkner, H.~Liu, J.~McGreevy, and D.~Vegh, {\it {Emergent quantum criticality, Fermi surfaces, and AdS(2)}},  {\em Phys. Rev. D} {\bf 83} (2011) 125002, [\href{http://arxiv.org/abs/0907.2694}{{\tt arXiv:0907.2694}}].

\bibitem{Liu:2009dm}
H.~Liu, J.~McGreevy, and D.~Vegh, {\it {Non-Fermi liquids from holography}},  {\em Phys. Rev. D} {\bf 83} (2011) 065029, [\href{http://arxiv.org/abs/0903.2477}{{\tt arXiv:0903.2477}}].

\bibitem{Lee:2008xf}
S.-S. Lee, {\it {A Non-Fermi Liquid from a Charged Black Hole: A Critical Fermi Ball}},  {\em Phys. Rev. D} {\bf 79} (2009) 086006, [\href{http://arxiv.org/abs/0809.3402}{{\tt arXiv:0809.3402}}].

\bibitem{Iqbal:2008by}
N.~Iqbal and H.~Liu, {\it {Universality of the hydrodynamic limit in AdS/CFT and the membrane paradigm}},  {\em Phys. Rev. D} {\bf 79} (2009) 025023, [\href{http://arxiv.org/abs/0809.3808}{{\tt arXiv:0809.3808}}].

\bibitem{Faulkner:2009am}
T.~Faulkner, G.~T. Horowitz, J.~McGreevy, M.~M. Roberts, and D.~Vegh, {\it {Photoemission 'experiments' on holographic superconductors}},  {\em JHEP} {\bf 03} (2010) 121, [\href{http://arxiv.org/abs/0911.3402}{{\tt arXiv:0911.3402}}].

\bibitem{Wu:2011cy}
J.-P. Wu, {\it {Some properties of the holographic fermions in an extremal charged dilatonic black hole}},  {\em Phys. Rev. D} {\bf 84} (2011) 064008, [\href{http://arxiv.org/abs/1108.6134}{{\tt arXiv:1108.6134}}].

\bibitem{Li:2011sh}
W.-J. Li, R.~Meyer, and H.-b. Zhang, {\it {Holographic non-relativistic fermionic fixed point by the charged dilatonic black hole}},  {\em JHEP} {\bf 01} (2012) 153, [\href{http://arxiv.org/abs/1111.3783}{{\tt arXiv:1111.3783}}].

\bibitem{Fang:2012pw}
L.~Q. Fang, X.-H. Ge, and X.-M. Kuang, {\it {Holographic fermions in charged Lifshitz theory}},  {\em Phys. Rev. D} {\bf 86} (2012) 105037, [\href{http://arxiv.org/abs/1201.3832}{{\tt arXiv:1201.3832}}].

\bibitem{Li:2012uua}
W.-J. Li and J.-P. Wu, {\it {Holographic fermions in charged dilaton black branes}},  {\em Nucl. Phys. B} {\bf 867} (2013) 810--826, [\href{http://arxiv.org/abs/1203.0674}{{\tt arXiv:1203.0674}}].

\bibitem{Chagnet:2022ykl}
N.~Chagnet, V.~Djuki{\'c}, M.~{\v{C}}ubrovi{\'c}, and K.~Schalm, {\it {Emerging Fermi liquids from regulated quantum electron stars}},  {\em JHEP} {\bf 08} (2022) 222, [\href{http://arxiv.org/abs/2204.10092}{{\tt arXiv:2204.10092}}].

\bibitem{Lu:2024qxj}
C.-Y. Lu, X.-H. Ge, and S.-J. Sin, {\it {Holographic fermions in the dyonic Gubser-Rocha black hole}},  {\em Phys. Rev. D} {\bf 111} (2025), no.~8 086011, [\href{http://arxiv.org/abs/2412.20160}{{\tt arXiv:2412.20160}}].

\bibitem{Lu:2025zxq}
C.-Y. Lu, X.-H. Ge, and S.-J. Sin, {\it {Lifshitz transition in a holographic finite density flavour brane Weyl semimetal}},  \href{http://arxiv.org/abs/2509.24287}{{\tt arXiv:2509.24287}}.

\bibitem{Liu:2012tr}
Y.~Liu, K.~Schalm, Y.-W. Sun, and J.~Zaanen, {\it Lattice potentials and fermions in holographic non fermi-liquids: Hybridizing local quantum criticality},  {\em JHEP} {\bf 10} (2012) 036, [\href{http://arxiv.org/abs/1205.5227}{{\tt arXiv:1205.5227}}].

\bibitem{Ling:2013aya}
Y.~Ling, C.~Niu, J.-P. Wu, Z.-Y. Xian, and H.-b. Zhang, {\it {Holographic Fermionic Liquid with Lattices}},  {\em JHEP} {\bf 07} (2013) 045, [\href{http://arxiv.org/abs/1304.2128}{{\tt arXiv:1304.2128}}].

\bibitem{Ling:2014bda}
Y.~Ling, P.~Liu, C.~Niu, J.-P. Wu, and Z.-Y. Xian, {\it {Holographic fermionic system with dipole coupling on Q-lattice}},  {\em JHEP} {\bf 12} (2014) 149, [\href{http://arxiv.org/abs/1410.7323}{{\tt arXiv:1410.7323}}].

\bibitem{Bagrov:2016cnr}
A.~Bagrov, N.~Kaplis, A.~Krikun, K.~Schalm, and J.~Zaanen, {\it {Holographic fermions at strong translational symmetry breaking: a Bianchi-VII case study}},  {\em JHEP} {\bf 11} (2016) 057, [\href{http://arxiv.org/abs/1608.03738}{{\tt arXiv:1608.03738}}].

\bibitem{Cremonini:2018xgj}
S.~Cremonini, L.~Li, and J.~Ren, {\it {Holographic Fermions in Striped Phases}},  {\em JHEP} {\bf 12} (2018) 080, [\href{http://arxiv.org/abs/1807.11730}{{\tt arXiv:1807.11730}}].

\bibitem{Cremonini:2019fzz}
S.~Cremonini, L.~Li, and J.~Ren, {\it {Spectral Weight Suppression and Fermi Arc-like Features with Strong Holographic Lattices}},  {\em JHEP} {\bf 09} (2019) 014, [\href{http://arxiv.org/abs/1906.02753}{{\tt arXiv:1906.02753}}].

\bibitem{Balm:2019dxk}
F.~Balm, A.~Krikun, A.~Romero-Berm{\'u}dez, K.~Schalm, and J.~Zaanen, {\it {Isolated zeros destroy Fermi surface in holographic models with a lattice}},  {\em JHEP} {\bf 01} (2020) 151, [\href{http://arxiv.org/abs/1909.09394}{{\tt arXiv:1909.09394}}].

\bibitem{Jeong:2019zab}
H.-S. Jeong, K.-Y. Kim, Y.~Seo, S.-J. Sin, and S.-Y. Wu, {\it {Holographic Spectral Functions with Momentum Relaxation}},  {\em Phys. Rev. D} {\bf 102} (2020), no.~2 026017, [\href{http://arxiv.org/abs/1910.11034}{{\tt arXiv:1910.11034}}].

\bibitem{Iliasov:2019pav}
A.~Iliasov, A.~A. Bagrov, M.~I. Katsnelson, and A.~Krikun, {\it {Anisotropic destruction of the Fermi surface in inhomogeneous holographic lattices}},  {\em JHEP} {\bf 01} (2020) 065, [\href{http://arxiv.org/abs/1910.01542}{{\tt arXiv:1910.01542}}].

\bibitem{Mukhopadhyay:2020tky}
S.~Mukhopadhyay and N.~Rai, {\it {Holographic Fermi surfaces in charge density wave from D2-D8}},  {\em JHEP} {\bf 09} (2021) 160, [\href{http://arxiv.org/abs/2008.00432}{{\tt arXiv:2008.00432}}].

\bibitem{Hercek:2022wyu}
F.~Her{\v{c}}ek, V.~Gecin, and M.~{\v{C}}ubrovi{\'c}, {\it {Photoemission ''experiments'' on holographic lattices}},  {\em SciPost Phys. Core} {\bf 6} (2023) 027, [\href{http://arxiv.org/abs/2208.05920}{{\tt arXiv:2208.05920}}].

\bibitem{Bednorz:1986tc}
J.~G. Bednorz and K.~A. Muller, {\it {Possible high Tc superconductivity in the Ba-La-Cu-O system}},  {\em Z. Phys. B} {\bf 64} (1986) 189--193.

\bibitem{Ling:2023ncu}
Y.~Ling, P.~Liu, and M.-H. Wu, {\it {The commensurate state and lock-in in a holographic model}},  {\em JHEP} {\bf 11} (2024) 143, [\href{http://arxiv.org/abs/2310.14324}{{\tt arXiv:2310.14324}}].

\bibitem{Kiritsis:2015hoa}
E.~Kiritsis and L.~Li, {\it {Holographic Competition of Phases and Superconductivity}},  {\em JHEP} {\bf 01} (2016) 147, [\href{http://arxiv.org/abs/1510.00020}{{\tt arXiv:1510.00020}}].

\bibitem{Benini:2010qc}
F.~Benini, C.~P. Herzog, and A.~Yarom, {\it {Holographic Fermi arcs and a d-wave gap}},  {\em Phys. Lett. B} {\bf 701} (2011) 626--629, [\href{http://arxiv.org/abs/1006.0731}{{\tt arXiv:1006.0731}}].

\bibitem{Kanigel}
A.~K. {\it et al.}, {\it {Evolution of the Pseudogap from Fermi Arcs to the Nodal Liquid}},  {\em Nature Physics} {\bf 02} (2006) 447.

\bibitem{Faulkner:2011tm}
T.~Faulkner, N.~Iqbal, H.~Liu, J.~McGreevy, and D.~Vegh, {\it {Holographic non-Fermi liquid fixed points}},  {\em Phil. Trans. Roy. Soc.} {\bf A 369} (2011) 1640, [\href{http://arxiv.org/abs/1101.0597}{{\tt arXiv:1101.0597}}].

\bibitem{boebinger1996insulator}
G.~Boebinger, Y.~Ando, A.~Passner, T.~Kimura, M.~Okuya, J.~Shimoyama, K.~Kishio, K.~Tamasaku, N.~Ichikawa, and S.~Uchida, {\it {Insulator-to-Metal Crossover in the Normal State of ${\mathrm{La}}_{2\ensuremath{-}\mathit{x}}{\mathrm{Sr}}_{\mathit{x}}{\mathrm{CuO}}_{4}$ Near Optimum Doping}},  {\em Phys. Rev. Lett.} {\bf 77} (1996) 5417.

\bibitem{Vojta01112009}
M.~Vojta, {\it Lattice symmetry breaking in cuprate superconductors: stripes, nematics, and superconductivity},  {\em Advances in Physics} {\bf 58} (2009), no.~6 699--820.

\bibitem{Edalati:2010ge}
M.~Edalati, R.~G. Leigh, K.~W. Lo, and P.~W. Phillips, {\it {Dynamical Gap and Cuprate-like Physics from Holography}},  {\em Phys. Rev. D} {\bf 83} (2011) 046012, [\href{http://arxiv.org/abs/1012.3751}{{\tt arXiv:1012.3751}}].

\bibitem{Edalati:2010ww}
M.~Edalati, R.~G. Leigh, and P.~W. Phillips, {\it {Dynamically Generated Mott Gap from Holography}},  {\em Phys. Rev. Lett.} {\bf 106} (2011) 091602, [\href{http://arxiv.org/abs/1010.3238}{{\tt arXiv:1010.3238}}].

\bibitem{Chakrabarti:2021qie}
S.~Chakrabarti, D.~Maity, and W.~Wahlang, {\it {Studying the holographic Fermi surface in the scalar induced anisotropic background}},  {\em Phys. Lett. B} {\bf 827} (2022) 136990, [\href{http://arxiv.org/abs/2108.10043}{{\tt arXiv:2108.10043}}].

\bibitem{Wahlang:2021oqk}
W.~Wahlang, {\it {Evolution of holographic Fermi surface from non-minimal couplings}},  {\em Eur. Phys. J. C} {\bf 82} (2022), no.~4 339, [\href{http://arxiv.org/abs/2112.05097}{{\tt arXiv:2112.05097}}].

\bibitem{Chakrabarti:2022lxl}
S.~Chakrabarti, D.~Maity, and W.~Wahlang, {\it {A note on the effects of magnetic field on holographic fermions with dipole-like coupling}},  \href{http://arxiv.org/abs/2204.06756}{{\tt arXiv:2204.06756}}.

\end{thebibliography}\endgroup

\end{document}